\documentclass[twocolumn,twocolappendix]{aastex63}

\usepackage{natbib,aas_macros,amsmath}
\usepackage{multirow,color}
\usepackage{natbib}
\usepackage{longtable}
\usepackage{lineno}

\def\farcs{%
 \mbox{%
  \kern  0.13ex.%
  \kern -0.95ex\arcsec%
  \kern -0.1ex%
 }%
}%

\newcommand{\ci}{[C\,{\sc i}]}
\newcommand{\oiii}{[O\,{\sc iii}]}
\newcommand{\oii}{[O\,{\sc ii}]}
\newcommand{\cii}{[C\,{\sc ii}]}
\newcommand{\nii}{[N\,{\sc ii}]}
\newcommand{\neiii}{[Ne\,{\sc iii}]}

\newcommand{\sii}{[S\,{\sc ii}]}
\newcommand{\siii}{[S\,{\sc iii}]}

\newcommand{\hst}{{\it HST}}
\newcommand{\jwst}{{\it JWST}}
\newcommand{\prospector}{\texttt{Prospector}}
\newcommand{\eazy}{\texttt{EAZY}}

\newcommand{\survey}{{DUALZ}}
\newcommand{\wide}{\textit{Wide}}
\newcommand{\deep}{\textit{Deep}}

\submitjournal{ApJS}
\shorttitle{
DUALZ -- ALMA $\times$ JWST Public Legacy Field Abell 2744
}
\shortauthors{Fujimoto et al.}

\begin{document}

\title{
DUALZ -- Deep UNCOVER-ALMA Legacy High-$Z$ Survey 
}

\correspondingauthor{Seiji Fujimoto}
\email{fujimoto@utexas.edu}
\author[0000-0001-7201-5066]{Seiji Fujimoto}\altaffiliation{Hubble Fellow}
\affiliation{
Department of Astronomy, The University of Texas at Austin, Austin, TX 78712, USA
}
\affiliation{
David A. Dunlap Department of Astronomy and Astrophysics, University of Toronto, 50 St. George Street, Toronto, Ontario, M5S 3H4, Canada
}
\affiliation{
Dunlap Institute for Astronomy and Astrophysics, 50 St. George Street, Toronto, Ontario, M5S 3H4, Canada
}

\author[0000-0001-5063-8254]{Rachel Bezanson}
\affiliation{Department of Physics and Astronomy and PITT PACC, University of Pittsburgh, Pittsburgh, PA 15260, USA}

\author[0000-0002-2057-5376]{Ivo Labbe}
\affiliation{Centre for Astrophysics and Supercomputing, Swinburne University of Technology, Melbourne, VIC 3122, Australia}

\author[0000-0003-2680-005X]{Gabriel Brammer}
\affiliation{Cosmic Dawn Center (DAWN), Niels Bohr Institute, University of Copenhagen, Jagtvej 128, K{\o}benhavn N, DK-2200, Denmark}

\author[0000-0002-0108-4176]{Sedona H. Price}
\affiliation{Department of Physics and Astronomy and PITT PACC, University of Pittsburgh, Pittsburgh, PA 15260, USA}

\author[0000-0001-9269-5046]{Bingjie Wang}
\affiliation{Department of Astronomy \& Astrophysics, The Pennsylvania State University, University Park, PA 16802, USA}
\affiliation{Institute for Computational \& Data Sciences, The Pennsylvania State University, University Park, PA 16802, USA}
\affiliation{Institute for Gravitation and the Cosmos, The Pennsylvania State University, University Park, PA 16802, USA}

\author[0000-0003-1614-196X]{John R. Weaver}
\affiliation{Department of Astronomy, University of Massachusetts, Amherst, MA 01003, USA}

\author[0000-0001-7440-8832]{Yoshinobu Fudamoto} 
\affiliation{Waseda Research Institute for Science and Engineering, Faculty of Science and Engineering, Waseda University, 3-4-1 Okubo, Shinjuku, Tokyo 169-8555, Japan}
\affiliation{National Astronomical Observatory of Japan, 2-21-1, Osawa, Mitaka, Tokyo, Japan}

\author[0000-0001-5851-6649]{Pascal A. Oesch}
\affiliation{Department of Astronomy, University of Geneva, Chemin Pegasi 51, 1290 Versoix, Switzerland}
\affiliation{Cosmic Dawn Center (DAWN), Niels Bohr Institute, University of Copenhagen, Jagtvej 128, K{\o}benhavn N, DK-2200, Denmark}

\author[0000-0003-2919-7495]{Christina C. Williams}
\affiliation{NSF’s National Optical-Infrared Astronomy Research Laboratory, 950 N. Cherry Avenue, Tucson, AZ 85719, USA}
\affiliation{Steward Observatory, University of Arizona, 933 North Cherry Avenue, Tucson, AZ 85721, USA}
\author[0000-0001-8460-1564]{Pratika Dayal}
\affiliation{Kapteyn Astronomical Institute, University of Groningen, 9700 AV Groningen, The Netherlands}

\author[0000-0002-1109-1919]{Robert Feldmann}
\affiliation{Institute for Computational Science, University of Zurich, Zurich, CH-8057, Switzerland}

\author[0000-0002-5612-3427]{Jenny E. Greene}
\affiliation{Department of Astrophysical Sciences, Princeton University, 4 Ivy Lane, Princeton, NJ 08544}

\author[0000-0001-6755-1315]{Joel Leja}
\affiliation{Department of Astronomy \& Astrophysics, The Pennsylvania State University, University Park, PA 16802, USA}
\affiliation{Institute for Computational \& Data Sciences, The Pennsylvania State University, University Park, PA 16802, USA}
\affiliation{Institute for Gravitation and the Cosmos, The Pennsylvania State University, University Park, PA 16802, USA}

\author[0000-0001-7160-3632]{Katherine E. Whitaker}
\affiliation{Department of Astronomy, University of Massachusetts, Amherst, MA 01003, USA}

\author[0000-0002-0350-4488]{Adi Zitrin}
\affiliation{Physics Department, Ben-Gurion University of the Negev, P.O. Box 653, Be\'er-Sheva 84105, Israel}
\author[0000-0002-7031-2865]{Sam E. Cutler}
\affiliation{Department of Astronomy, University of Massachusetts, Amherst, MA 01003, USA}

\author[0000-0001-6278-032X]{Lukas J. Furtak}
\affiliation{Physics Department, Ben-Gurion University of the Negev, P.O. Box 653, Be\'er-Sheva 84105, Israel}

\author[0000-0002-9651-5716]{Richard Pan}
\affiliation{Department of Physics and Astronomy, Tufts University, 574 Boston Ave., Medford, MA 02155, USA}

\author[0009-0009-9795-6167]{Iryna Chemerynska}
\affiliation{Institut d'Astrophysique de Paris, CNRS, Sorbonne Universit\'e, 98bis Boulevard Arago, 75014, Paris, France}

\author[0000-0002-5588-9156]{Vasily Kokorev}
\affiliation{Kapteyn Astronomical Institute, University of Groningen, 9700 AV Groningen, The Netherlands}

\author[0000-0001-8367-6265]{Tim B. Miller}
\affiliation{Department of Astronomy, Yale University, New Haven, CT 06511, USA}
\affiliation{Center for Interdisciplinary Exploration and Research in Astrophysics (CIERA) and
Department of Physics and Astronomy, Northwestern University, 1800 Sherman Ave, Evanston IL 60201, USA}

\author[0000-0002-7570-0824]{Hakim Atek}
\affiliation{Institut d'Astrophysique de Paris, CNRS, Sorbonne Universit\'e, 98bis Boulevard Arago, 75014, Paris, France}

\author[0000-0002-8282-9888]{Pieter van Dokkum}
\affiliation{Department of Astronomy, Yale University, New Haven, CT 06511, USA}

\author[0000-0002-0000-2394]{St\'ephanie Juneau}
\affiliation{NSF’s National Optical-Infrared Astronomy Research Laboratory, 950 N. Cherry Avenue, Tucson, AZ 85719, USA}

\author[0000-0002-3838-8093]{Susan Kassin}
\affiliation{Space Telescope Science Institute (STScI), 3700 San Martin Drive, Baltimore, MD 21218, USA}

\author[0000-0002-3475-7648]{Gourav Khullar}
\affiliation{Department of Physics and Astronomy and PITT PACC, University of Pittsburgh, Pittsburgh, PA 15260, USA}

\author[0000-0001-9002-3502]{Danilo Marchesini}
\affiliation{Physics and Astronomy Department, Tufts University, 574 Boston Ave., Medford, MA 02155, USA}

\author[0000-0003-0695-4414]{Michael Maseda}
\affiliation{Department of Astronomy, University of Wisconsin-Madison, 475 N. Charter St., Madison, WI 53706 USA}

\author[0000-0002-7524-374X]{Erica J. Nelson}
\affiliation{Department for Astrophysical and Planetary Science, University of Colorado, Boulder, CO 80309, USA}

\author[0000-0003-4075-7393]{David J. Setton}
\affiliation{Department of Physics and Astronomy and PITT PACC, University of Pittsburgh, Pittsburgh, PA 15260, USA}

\author[0000-0001-8034-7802]{Renske Smit}
\affiliation{Astrophysics Research Institute, Liverpool John Moores University, 146 Brownlow Hill, Liverpool L3 5RF, UK}

\def\apj{ApJ}%
\def\apjl{ApJ}%
\def\apjs{ApJS}%

\def\rme{\rm e}
\def\rmstar{\rm star}
\def\rmFIR{\rm FIR}
\def\itHubble{\it Hubble}
\def\rmyr{\rm yr}

\begin{abstract}
We present the survey design and initial results of the ALMA Cycle~9 program of \survey, which aims to establish a joint ALMA and \jwst\ public legacy field targeting the massive galaxy cluster Abell~2744. \survey\ features a contiguous $4'\times6'$ ALMA 30-GHz-wide mosaic in Band~6, covering areas of $\mu>2$ down to a sensitivity of $\sigma=32.7$~$\mu$Jy.
Through a blind search, we identified 69 dust continuum sources at S/N $\gtrsim5.0$ with median redshift and intrinsic 1.2-mm flux of $z=2.30$ and $S_{\rm 1.2mm}^{\rm int}=0.24$~mJy. 
Of these, 27 have been spectroscopically confirmed, leveraged by the latest NIRSpec observations, while photometric redshifts are also constrained by the comprehensive \hst, NIRCam, and ALMA data sets for the remaining sources.
With priors, we further identify a \cii158~$\mu$m line emitter at $z=6.3254\pm0.0004$, 
confirmed by the latest NIRSpec spectroscopy. 
The NIRCam counterparts of the 1.2-mm continuum exhibit undisturbed morphologies, denoted either by disk or spheroid, 
implying the triggers for the faint mm emission are less catastrophic than mergers.
We have identified 8 \hst-dark galaxies (F150W$>$27~mag, F150W$-$F444W$>$2.3) and 2 \jwst-dark (F444W$>$30~mag) galaxy candidates among the ALMA continuum sources. 
The former includes face-on disk galaxies, hinting that substantial dust obscuration does not always result from inclination. 
We also detect a marginal dust emission from an X-ray-detected galaxy at $z_{\rm spec}=10.07$, suggesting an active co-evolution of the central black hole and its host. We assess the infrared luminosity function up to $z\sim10$
and find it consistent with predictions from galaxy formation models.
To foster diverse scientific outcomes from the community, we publicly release reduced ALMA mosaic maps, cubes, and the source catalog\footnote{\url{https://jwst-uncover.github.io/DR2.html\#DUALZ}}.
\end{abstract}

\section{Introduction}
\label{sec:intro} 

Star-forming activity plays an important role in the mass assembly of galaxies. 
A large amount of dust through supernovae (SNe) and asymptotic giant branch (AGB) stars is also produced in this process, which can make the activity dust obscured. 
Because of the thermal power of dust heated by these intense star-forming activities re-emitting in infrared (IR) wavelengths and of the negative-$k$ correction at millimeter (mm) and sub-millimeter (submm), IR observations, especially at mm and submm, are powerful probes to comprehensively study the formation and evolution of galaxies (see reviews e.g., \citealt{casey2014, hodge2020}). 

Previous IR observations using single-dish telescopes have revealed the presence of a unique high-redshift ($z>1$) population that is bright at submm--mm wavelengths, with extremely high star-formation rates (SFRs) reaching up to the order of $\sim$1000 $M_{\odot}$/yr \citep[SMGs; e.g., ][]{smail1997,hughes1998}. However, the number density of SMGs is relatively small and contributes only around $\sim$10--20\% to the cosmic infrared background light \citep[e.g.,][]{eales2000,smail2002,coppin2006,knudsen2008,weiss2009,perera2008,hatsukade2011,scott2012, cowie2017}, suggesting that more abundant faint mm and submm populations ($S_{\rm 1mm}\lesssim$~1~mJy) dominate the majority of the dust-obscured mass assembly of galaxies throughout cosmic history.

The Atacama Large Millimeter/submillimeter Array (ALMA) provides a unique FIR window to identify and characterize these faint submm and mm populations, thanks to its improved sensitivity and angular resolution \citep[e.g.,][]{hatsukade2013,ono2014,carniani2015,fujimoto2016, oteo2016, aravena2016, walter2016, dunlop2017,umehata2017,zavala2018,hatsukade2018,franco2018, arancibia2019, gonzalez2020, bethermin2020, klitsch2020, zavala2021, gomez2021, cowie2022, fujimoto2023, cowie2023}.
These studies indicate that the faint submm and mm sources newly identified with ALMA are more numerous than the SMGs, contributing to the CIB by $\sim$70$-$100\% down to $\sim$0.01 mJy \citep[e.g.,][]{arancibia2019, fujimoto2023}, and these deep ALMA observations have succeeded in detecting faint dust emission even from main-sequence galaxies \citep[e.g.,][]{aravena2020}.
These new ALMA sources are typically characterized as massive galaxies with the stellar mass of $M_{\rm star}\gtrsim10^{10}M_{\odot}$ at $z\simeq$ 1--3 \citep[e.g.,][]{dunlop2017, aravena2020}, while potentially less massive ($M_{\rm star}\lesssim10^{10}M_{\odot}$) dusty galaxies are also identified even at $z>7$ \citep{fudamoto2021}. 
Successful dust continuum detection has also been made from follow-up ALMA observations for UV-selected galaxies at $z\simeq$ 4.5--7.5 \citep[e.g.,][]{bethermin2020,bouwens2021}. 
However, these ALMA observations also reveal that the majority of UV-selected galaxies do not display detectable dust emission \citep[e.g.,][]{bouwens2020}, and the triggers for dusty star-forming activities remain unclear.

The advent of \jwst\ offers an unparalleled opportunity to characterize distant galaxies in exceptional detail.
The superb spatial resolution of NIRCam is almost three times better than that of {\it HST} \citep{gardner2023}, and its new spectroscopic window at the NIR--MIR wavelengths can detect a rich variety of nebular emission lines and enable unprecedented ISM characterizations \citep[e.g.,][]{pontoppidan2022}.
This opportunity paves the way for an in-depth characterization of dusty galaxies, such as their internal structures, color gradients, kinematics, chemical enrichment, and ionization state \citep[e.g.,][]{casey2017, suzuki2021, tchen2022, rujopakarn2023, clara2023, kokorev2023}, thereby complementing and enhancing our understanding derived from ALMA observations.

In this paper, we present the Deep UNCOVER-ALMA Legacy High-$z$ (DUALZ) Survey, specifically designed to establish Abell 2744 (A2744), one of the best-studied massive galaxy lensing clusters, as a joint ALMA and \jwst\ legacy field.
This is the first public ALMA survey that uniquely positions us at the intersection of both ALMA and \jwst\ data, aiming to support a broad array of legacy science from the community. Consequently, we make both ALMA and \jwst\ data publicly available\footnote{\url{https://jwst-uncover.github.io/}}.
In Section 2, we describe the survey design, observations, and data processing of \survey.
Section 3 outlines the methods for ALMA source extraction, identification of NIRCam counterparts, and the derivation of basic physical properties.
In Section 4, we present the initial outcomes of this survey, and in Section 5, we overview several examples of potential further science cases.
We summarize this study in Section 6.
Throughout this paper, we assume a flat universe with
$\Omega_{\rm m} = 0.3$,
$\Omega_\Lambda = 0.7$,
$\sigma_8 = 0.8$,
and $H_0 = 70$ km s$^{-1}$ Mpc$^{-1}$.
We use magnitudes in the AB system \citep{oke1983}.
We account for the cosmic microwave background (CMB) effect following the recipe presented by \cite{dacunha2013} (see also \citealt{pallottini2015, zhang2016, lagache2018}).
\section{Observations} 
\label{sec:data}

\begin{table*}
\setlength{\tabcolsep}{6pt}
\begin{center}
\caption{Log of ALMA observations in A2744}
\vspace{-0.4cm}
\label{tab:obslog}
\begin{tabular}{cccccccc}
\hline 
\hline
Program ID & UT start date  & $L_{\rm base}$ [m] & $N_{\rm ant}$ & Area &  Tuning & $t_{\rm obs}$[min] & PWV[mm]    \\ 
(1)        &  (2)           &  (3)           &     (4)       &  (5)    & (6)  & (7)           & (8)    \\ \hline
2022.1.00073.S & 2022-10-04, 10-10 & 15--500  & 43              &   A     & T1   &  133.0       &  0.60 \\
               & 2022-10-14     & 15--500  & 44              &   A     & T2   &  132.9       &  0.71 \\
               & 2022-10-07, 10-11  & 15--500  & 43             &   A      & T3   &  133.2       & 0.56 \\
               & 2022-10-17     & 14--456  & 45              &   A      & T4   &  133.5       &  1.39 \\
               & 2022-10-21     & 14--368  & 43              &   B      & T1     &  133.1      & 0.40 \\
               & 2022-10-20, 10-21 & 14--368  & 43              &   B      & T2   &  133.0       & 0.49 \\
               & 2022-10-07, 10-10 & 15--500  & 45              &   B      & T3     &  133.1      & 0.42 \\
               & 2022-10-18     & 14--456  & 45              &   B      & T4    &  133.9       &  1.04 \\
               & 2022-10-16, 10-17 & 14--456  & 45              &   C     & T1     &  133.1      & 1.20  \\
               & 2022-10-14, 10-16 & 14--483  & 43              &   C      & T2     &  134.2      & 0.57 \\
               & 2022-10-03     & 15--500  & 43              &   C      & T3     &  132.8      &  0.46 \\
               & 2022-10-17     & 14--456  & 45              &   C     & T4    &  134.8      & 1.10 \\
               & 2022-10-07, 10-10 & 15--500  & 43              &   D     & T1     &  133.3      &  0.42 \\
               & 2022-10-19, 10-20 & 14--368  & 45              &   D      & T2     &  132.9      & 0.70 \\
               & 2022-10-21, 10-22 & 14--368  & 44              &   D      & T3     &  133.3      & 0.42 \\
               & 2022-10-04, 10-06 & 15--500  & 43              &   D     & T4    &  133.2      & 0.55 \\ \hline 
2018.1.00035.L & 2019-03-14, 03-16 & 14--360  & 45              &   primary cluster    & T5    &  90.5   &  0.55 \\ \hline  
2013.1.00999.S & 2014-06-29, 07-29, 12-24, 12-31 & 14--820  & 33   &   primary cluster  & T6  &  354.6 &  1.06    \\
\hline
\end{tabular}
\end{center}
\vspace{-0.4cm}
\tablecomments{
(1) ALMA project ID. 
(2) Observation starting date in UTC (YYYY-MM-DDDD).  
(3) Baseline length. 
(4) Average number of 12-m antenna used for the observations. 
(5) Observing area. In \#2022.1.00073.S, the observations were split into four tiles of A, B, C, and D to fully cover the UNCOVER area of $\sim4'\times6'$.   
(6) Frequency tuning ID (T1: 244.01--247.75~GHz \& 259.01--262.75~GHz , T2: 247.76--251.50~GHz \& 262.76--266.50~GHz, T3: 251.51--255.25~GHz \& 266.51--270.25~GHz, T4: 255.26--259.00~GHz \& 270.26--274.00~GHz, T5: 253.8--257.6~GHz \& 268.8--272.6, T6: 250.1--253.8~GHz \& 265.1--268.8~GHz). 
(7) Observing time, including overheads and calibrations. 
(8) Average Precipitable Water Vapor (PWV) during observations. 
}
\end{table*}

\begin{figure*}
\begin{center}
\includegraphics[trim=0cm 0cm 0cm 0cm, clip, angle=0,width=0.8\textwidth]{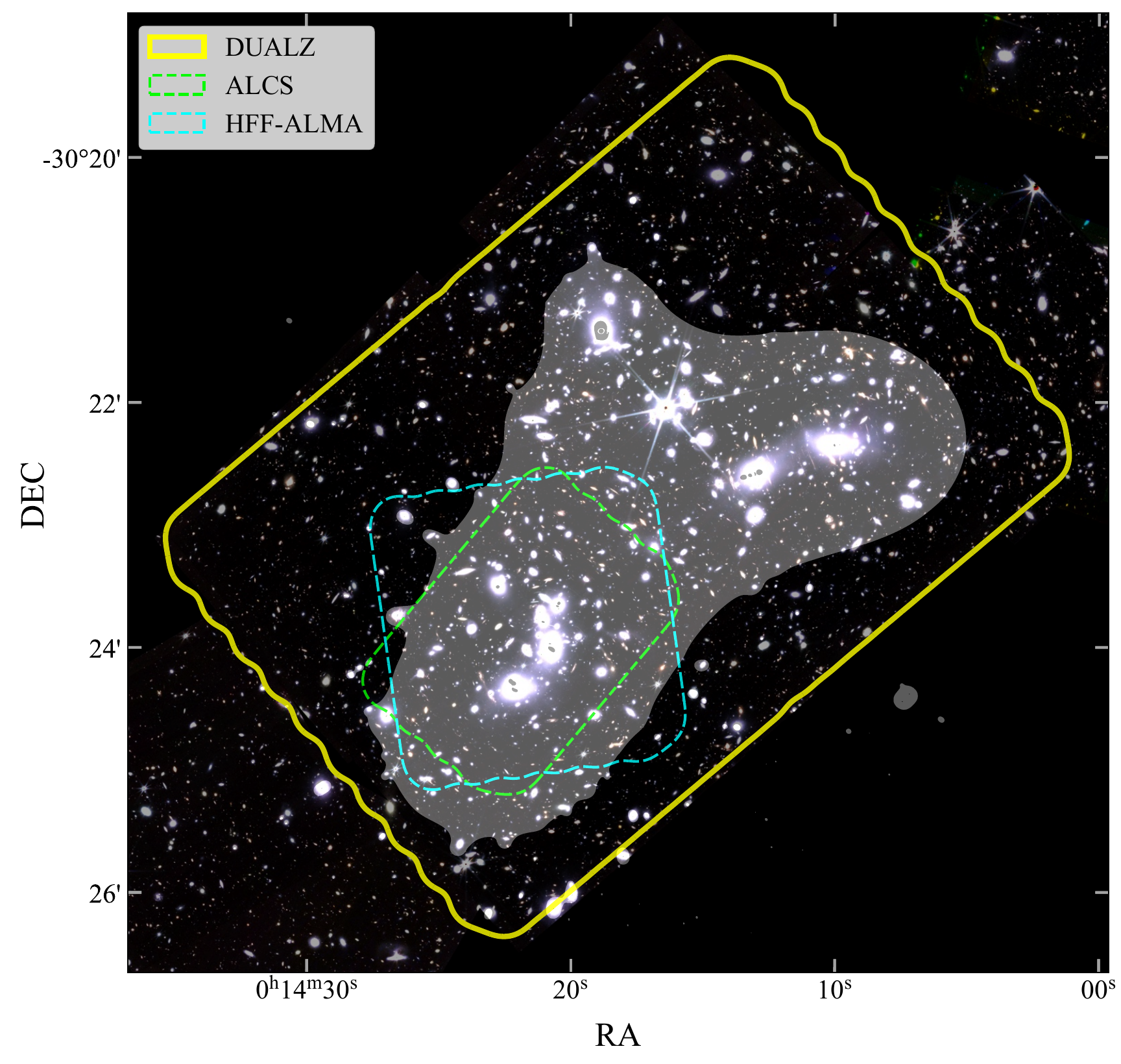}
\end{center}
\vspace{-0.6cm}
 \caption{
ALMA footprints overlaid on the NIRCam RBG (R: F444W, G: F356W, B: F277W) map of A2744 taken in UNCOVER \citep{bezanson2022}. 
The yellow solid curve shows the relative sensitivity response to the deepest 20\% of the mosaic of \survey, matched to the NIRCam footprint of the UNCOVER survey, 
and the green and cyan dashed curves present those of ALMA-HFF \citep{arancibia2023} and ALCS \citep{fujimoto2023b}. 
The white-shaded region indicates the highly magnified area with magnifications of $\geq2$ \citep{furtak2023a}.  
\label{fig:footprint}}
\end{figure*}

\begin{figure*}[t!]
\includegraphics[trim=0cm 0cm 0cm 0cm, clip, angle=0, width=1.0\textwidth]{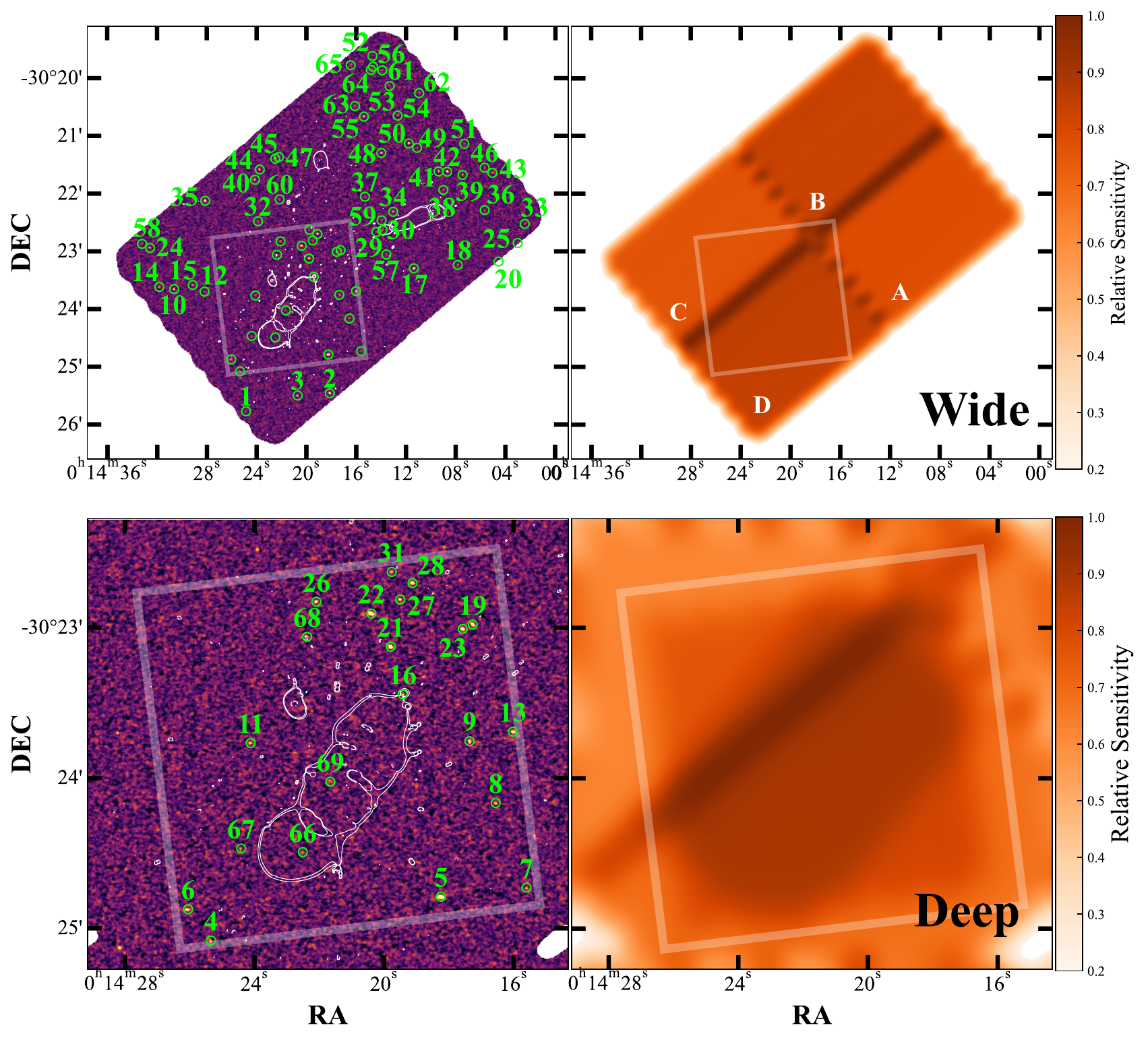}
 \caption{
ALMA Band~6 continuum mosaic maps without primary beam correction (left) and relative sensitivity response maps of the mosaic (right) of \survey-wide (top panels) and \survey-deep (bottom panels).
The \survey-wide mosaic was accomplished by combining the four mosaics of A, B, C, and D labeled on the top right panel.  
The labeled green circles represent the ALMA continuum sources.  
The white lines denote the $\mu=200$ magnification curve at $z=6$. 
The white square indicates the footprint of the ALMA-HFF, inside of where we use the \survey-deep map to complementary identify further faint ALMA sources. 
\label{fig:alma_mosaic}}
\end{figure*}

\subsection{Survey Design}
\label{sec:survey}

Abell 2744 (A2744), at $z=0.308$, is one of the best-studied massive galaxy clusters and is the target of our survey. A2744 has been extensively observed using the \textit{Hubble Space Telescope} (\hst) as part of the Hubble Frontier Field Survey (HFF; \citealt{lotz2017}).
A2744 features a low infrared background and a high magnification area that aligns well with the NIRCam field of view. Over 200 hours of JWST Cycle 1 and Cycle 2 observations, including GTO, ERS, GO, and DDT programs, have been performed and further scheduled toward this cluster.
In particular, the 82-hour public GO treasury program in \jwst\ Cycle 1 -- UNCOVER (\#2561; PIs I. Labbe \& R. Bezanson) is designed to obtain deep NIRCam and NIRSpec observations over an extended $4'\times6'$ field, covering the area with magnifications of $\mu\geq2$ around the primary cluster observed in HFF and two additional subclusters at northern and western regions \citep{furtak2023a}. UNCOVER consists of two components: 1) a deep NIRCam pre-imaging mosaic in 8 filters for 3.7--6.0 hours per band in late 2022 and 2) an ultra-deep 2.7--17.4~hours NIRSpec/prism low-resolution follow-up of NIRCam-detected high-redshift galaxies roughly 6 months later \citep{bezanson2022}.

The ALMA Band 6 program of \survey\ was designed to homogeneously map the main UNCOVER $4'\times6'$ field at the 1.2-mm wavelength and was accepted in Cycle 9 (\#2022.1.00073.S; PI: S. Fujimoto).
The full $4'\times6'$ mosaic map was achieved with four tiles due to the maximum pointing number limitation in ALMA mosaic observations.
Four frequency setups were used to carry out a 30-GHz wide spectral scan from 244 GHz to 274 GHz to reach the requested continuum sensitivity, thereby maximizing the chances of serendipitous line detection.
Figure \ref{fig:footprint} shows the ALMA footprint of \survey\ overlaid on the NIRCam color image obtained in A2744.
We also illustrate the ALMA footprints of two other ALMA Band 6 programs, ALMA-HFF \citep{arancibia2023} and ALMA Lensing Cluster Survey\footnote{\url{http://www.ioa.s.u-tokyo.ac.jp/ALCS/}} (ALCS; \citealt{fujimoto2023}), designed to observe the primary cluster observed in HFF.
Compared to previous ALMA programs, \survey\ increases the homogeneous 1.2-mm mapping area in A2744 by approximately a factor of 6.

\subsection{Observation and Data}
\label{sec:alma_reduction}

The ALMA Band 6 observations were completed in October 2022. 
Although a cyberattack on the ALMA observatory caused a delay in the QA process, the data was not impacted by the incident.
The observations were conducted under the array configurations of C-2 and C-3 with 43-45 antennae, baseline lengths ranging from 14-368 m to 15-500 m, and average precipitable water vapor (PWV) of 0.42-1.20 mm. The total observation time was 33.3 hours.
Bandpass calibrations were performed using J2258-2758 and J0334-4008, while phase calibrations were carried out using J2359-3133.
The observation log is summarized in Table~\ref{tab:obslog}, along with that of the previous ALMA programs.

The ALMA data were processed and calibrated with the Common Astronomy Software Applications package version 6.4.12 (CASA; \citealt{casa2022}) using the standard pipeline script.
We generated images from the calibrated visibilities using the natural weighting, a pixel scale of $0\farcs10$, and a primary beam limit down to 0.1, by running the CASA task {\sc tclean}.
For the continuum maps, the {\sc tclean} routines were executed down to the 2$\sigma$ level with a maximum iteration number of 100,000 in the automask mode\footnote{
We adopt {\sc noisethreshold=4.25}, {\sc sidelobethreshold=2.0}, {\sc minbeamfrac=0.3}, {\sc lownoisethreshold=1.5}, {\sc negativethreshold=0.0}, following the CASA automasking guide.
}.
For the cubes, we adopted a common spectral channel bin of 50 km~s$^{-1}$.
As we did not find any significant signals standing out in each channel either by bright line and/or continuum emitters in the cube with the spectral resolution above, we used the cube produced without the CLEAN iteration.
To avoid missing any strongly lensed (distorted) objects, we also produced a lower-resolution map and cube by applying a $uv$-taper parameter of $1\farcs5\times1\farcs5$.
We refer to our ALMA maps (cubes) without and with the $uv$-taper as natural and tapered maps (cubes), respectively.

\begin{figure}
\includegraphics[trim=0cm 0cm 0cm 0cm, clip, angle=0,width=0.48\textwidth]{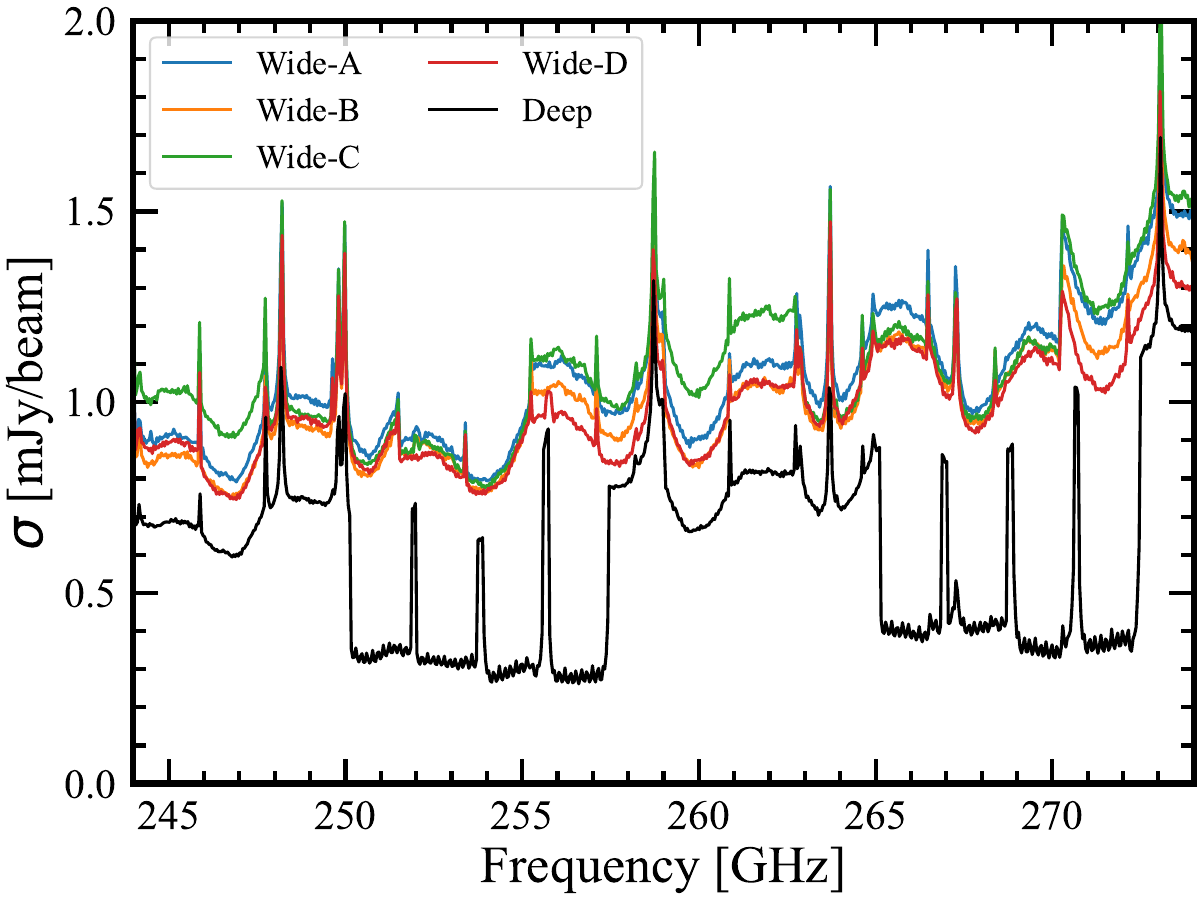}
\vspace{-0.4cm}
 \caption{Sensitivity (PB=1.0) of the \wide-natural and \deep-natural data cube. 
 The previous data covers 250--257.5~GHz and 265--272.5~GHz, where the sensitivity of the deep cube is better than that of the \wide\ cube by a factor of $\sim3$.  
\label{fig:line_sensitivity}}
\end{figure}

To leverage the ancillary data sets from the previous ALMA programs of ALMA-HFF and ALCS, we also created a deep ALMA map around the primary cluster.
The previous ALMA data were reduced, calibrated, and combined in the same manner as \cite{fujimoto2023b}.
We further combined\footnote{
We use CASA task {\sc concat}, where the weight is applied based on the data depths.}
 our calibrated visibilities of \survey\ for the pointings that fall within a $20\farcs0$ radius from the footprints of ALMA-HFF and ALCS. 
The natural and tapered maps and cubes were produced in the same manner as above.
We refer to the ALMA maps from the \survey\ data and this combined data set around the primary cluster as \wide\ and \deep\ maps, respectively.
In total, we produced 4 types of ALMA maps (cubes) in \survey\ -- \wide-natural, \wide-tapered, \deep-natural, and \deep-tapered maps (cubes).
In Figure~\ref{fig:alma_mosaic}, we show the \wide-natural and \deep-tapered maps and their relative sensitivity response maps.
The basic data properties (depth, beam size) are summarized in Table~\ref{tab:data_prop}.
The sensitivity as a function of frequency in our data cubes is also summarized in Figure~\ref{fig:line_sensitivity}.

\begin{table}[h]
\setlength{\tabcolsep}{12pt}
\begin{center}
\caption{Properties of ALMA maps}
\vspace{-0.4cm}
\label{tab:data_prop}
\begin{tabular}{ccc}
\hline 
\hline
Map & $\sigma$ [$\mu$Jy~beam$^{-1}$]  & beam [arcsec]\\ 
(1) &  (2)                            &  (3)   \\ \hline
\multicolumn{3}{c}{\wide\ ($\sim$$4'\times6'$)} \\ \hline
    \wide-natural    & 32.7 &  $1.02\times0.77$    \\
    \wide-tapered    & 43.6 & $1.81\times1.60$     \\ \hline
\multicolumn{3}{c}{\deep\ ($\sim$$2'\times2'$)} \\ \hline
    \deep-natural    & 27.7 & $1.00\times0.75$  \\
    \deep-tapered    & 38.6 & $1.81\times1.60$  \\  
    \hline
\end{tabular}
\end{center}
\vspace{-0.4cm}
\tablecomments{
(1) Map names.  
(2) Sensitivity (PB=1.0) of the continuum maps evaluated by the standard deviation of the pixel count. 
(3) Full-width-half-maximum (FWHM) of the synthesized beam size. 
}
\end{table}

\subsection{Survey Area}

\begin{figure}
\includegraphics[trim=0cm 0cm 0cm 0cm, clip, angle=0,width=0.48\textwidth]{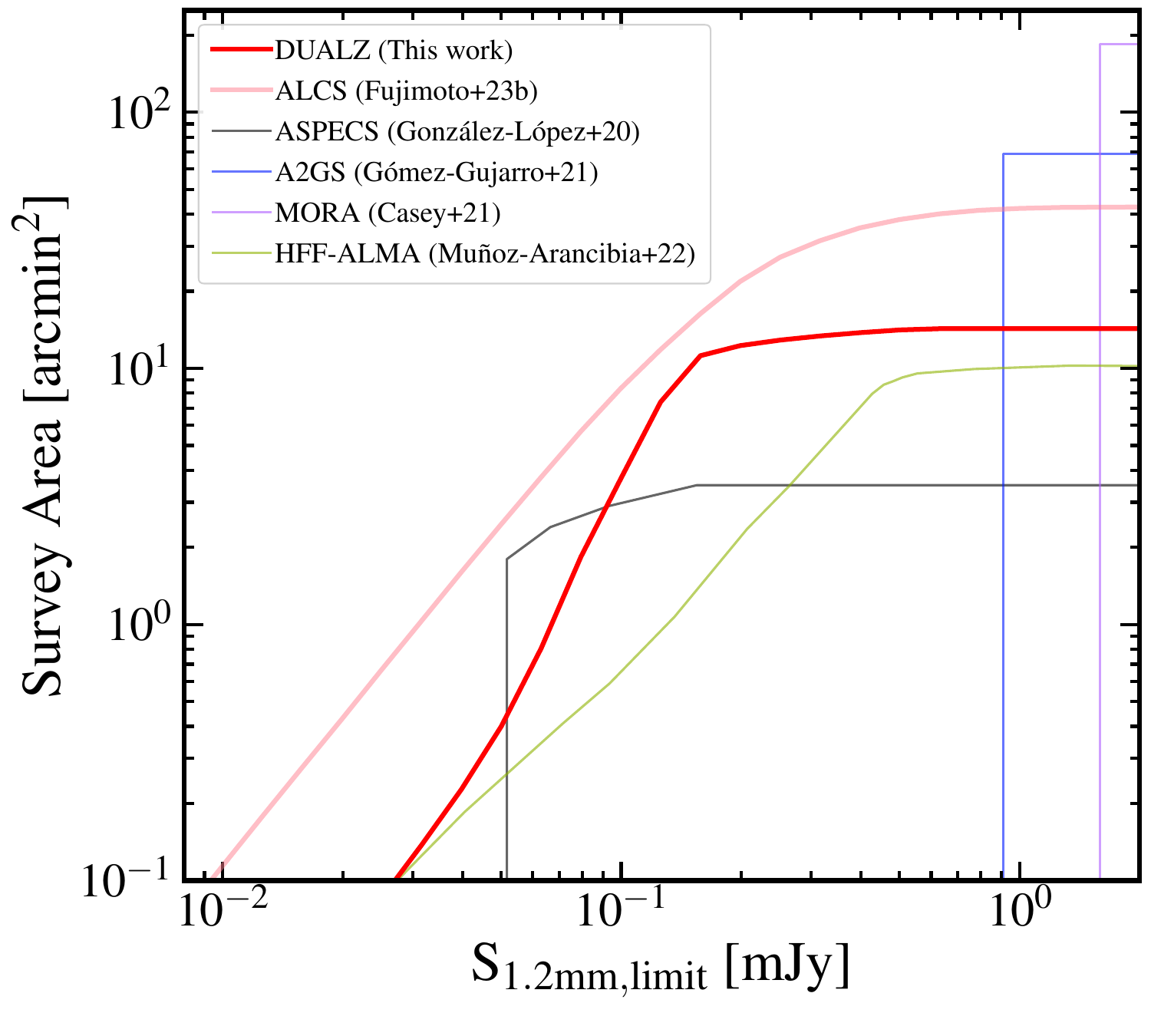}
\vspace{-0.4cm}
 \caption{
Survey area against the limiting 1.2-mm flux ($S_{\rm 1.2mm, limit}$; $5\sigma$) for \survey\ and other large ALMA surveys in the literature \citep{gonzalez2020, gomez2021, casey2021, arancibia2023, fujimoto2023b}.
For the lensing cluster studies, the lens correction is applied assuming $z=2.0$.
\label{fig:survey_area}}
\end{figure}

Figure~\ref{fig:survey_area} shows the effective survey area of \survey\ after applying the lensing correction at $z=2$, using the lens model presented in \cite{furtak2023a}. 
At a given limiting source flux at 1.2mm ($S_{\rm 1.2mm, limit}$), we calculate the areas whose lens-corrected sensitivities detect the source flux at levels $\geq5\sigma$, down to the relative response to the deepest $\geq$20\% of the \wide\ mosaic map.
For comparison, Figure \ref{fig:survey_area} also illustrates the effective survey areas estimated in the same manner as above, applying the lens correction when necessary, from the recent ALMA surveys.
This figure demonstrates the capacity of \survey\ to efficiently explore the faint ($\lesssim0.5$~mJy) regime more extensively ($\gtrsim3$~arcmin$^{2}$) than most other ALMA surveys.

\section{Data Analysis}
\label{sec:analysis}

\subsection{ALMA Source Extraction}
\label{sec:source_ext}

\begin{figure*}
\includegraphics[trim=0cm 0cm 0cm 0cm, clip, angle=0,width=1.\textwidth]{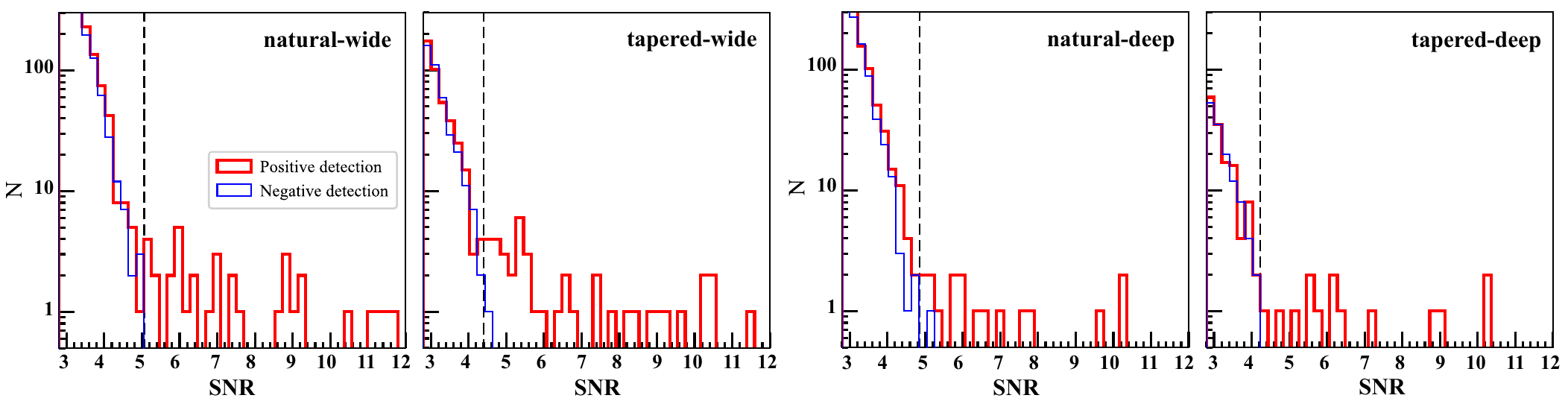}
\caption{
Differential counts of positive (red) and negative (blue) sources as a function of peak SNR.
From left to right, we present the results obtained from the \wide-natural, \wide-tapered, \deep-natural, and \deep-tapered maps. 
The vertical dashed line denotes the SNR threshold adopted in this paper for each map, to reliably identify ALMA sources in a blind manner.
\label{fig:pn_hist}}
\end{figure*}

We conduct the ALMA source extraction following the same procedure as \cite{fujimoto2023b} for the \wide-natural and \wide-tapered maps across the $\sim 4’\times6'$ area.
We use the maps before primary beam (PB) correction for source extraction.
We run version 2.5.0 of {\sc sextractor} \citep{bertin1996} and extract sources showing positive peak counts exceeding a 2.0$\sigma$ threshold, where an island of emission is considered a single source.
To estimate the expected number of genuine sources, we also implement a negative peak analysis \citep[e.g.,][]{hatsukade2013,ono2014,carniani2015,fujimoto2016, fujimoto2023}.
We produce inverted maps by multiplying both natural and tapered maps by $-1$.
Following this, {\sc sextractor} is applied again to extract sources displaying negative peak counts exceeding the 2.0$\sigma$ level. In both positive and negative maps, we use sources that are identified in regions whose relative sensitivity to the deepest part of the mosaic extends down to 20\%.

In Figure \ref{fig:pn_hist}, we present histograms of the positive and negative sources as a function of the signal-to-noise ratio (SNR) at the peak count, based on the sources selected from the above-described procedure. The excess of positive sources over negative sources in these histograms suggests the existence of real sources.
We find a clear excess extending down to an SNR of $\gtrsim5$ ($\gtrsim4$) in the natural (tapered) maps. We also observe the excess likely continuing down to SNR$\sim$3.5 in every map. In this paper, our goal is to present secure sources as initial results and thus adopt the SNR thresholds at 5.0 and 4.4 in the \wide-natural and -tapered maps, respectively.

To complementarily detect even fainter sources from the \deep\ maps, we also apply the same procedures to the deep-natural and tapered maps. Based on the positive and negative source histograms, we set the SNR thresholds at 4.8 and 4.2 for the deep-natural and -tapered maps, respectively.

With these SNR thresholds, we identify 65 ALMA sources from the \wide\ maps and an additional 4 ALMA sources from the deep maps.
We summarize the total of 69 ALMA sources and their ALMA ID (AID) in Table~\ref{tab:catalog}.
Of note, Figure~\ref{fig:pn_hist} shows that a single negative source exceeding the SNR thresholds remains in both \wide\ and \deep\ maps. This implies that $\sim$1--2 out of the 69 sources might be spurious, while also demonstrating the high purity $p>0.97$ of our ALMA sources \citep[e.g.,][]{gonzalez2020, gomez2021}, defined as
\begin{equation}
p = \frac{N_{\rm pos}-N_{\rm neg}}{N_{\rm pos}},
\end{equation}
where $N_{\rm pos}$ and $N_{\rm neg}$ represent the number of positive and negative sources at a given SNR, respectively.
We further discuss the potential of spurious sources in Section \ref{sec:jwst-dark}.

The 1.2-mm flux densities for our 69 ALMA sources are measured using three different methods: 1) peak pixel count, 2) $1\farcs0$-radius aperture photometry, and 3) 2D elliptical Gaussian fitting with CASA {\sc imfit}. We also assess the SNR of each measurement. If the SNR of the {\sc imfit} measurement exceeds 10, we adopt the {\sc imfit} measurement; otherwise, we use the measurement with the best SNR. These measurements are performed with the maps prior to PB correction, while the PB correction is subsequently applied to the output values. In Table~\ref{tab:catalog}, we also list the final 1.2-mm flux measurement after the PB correction.

\subsection{NIRCam Counterparts}
\label{sec:counterpart}

We cross-match our ALMA continuum source catalog with the DR2 NIRCam source catalog presented in \cite{weaver2023}.
The source extraction of the NIRCam catalog was performed with a combined image of F277W+F356W+F444W, and the photometry was processed with all public \jwst/NIRCam, HST/ACS, and HST/WFC3 imaging available in A2744, including the \jwst\ Cycle~1 programs of UNCOVER (\#2561; PIs I.~Labbe \& R.~Bezanson; \citealt{bezanson2022}), GLASS (\# 1324; PI: T.~Treu; \citealt{treu2022}), and DDT (\#2756; PI: W.Chen; \citealt{roberts-borsani2023}). These NIRCam observations homogeneously observed the wavelength range at $\sim1$--5$\mu$m with 8 filters of F090W, F115W, F150W, F200W, F277W, F356W, F410M, and F444W down to the 5$\sigma$ limiting magnitudes of $\sim$29--30mag for point sources \citep{bezanson2022}.

With a search radius of $1\farcs0$, we identify 67 ALMA sources with NIRCam counterparts. 
When we identify multiple NIRCam sources within the search radius, we adopt the nearest one as the counterpart. 
The spatial offset ranges from $0\farcs0$--$0\farcs7$ with a median of $0\farcs12$, close to the pixel scale of $0\farcs1$ in our ALMA maps (Section~\ref{sec:alma_reduction}).
The high fraction (67/69) of the NIRCam presence and the small median spatial offset validate the robust selection of our ALMA continuum sources and the decent NIRCam sensitivity, even for dusty objects.
For the remaining two sources without NIRCam counterparts, AID20 falls slightly outside the edge of the NIRCam footprint, while AID60 does not show any counterparts in the NIRCam filters, even in F444W. This suggests that the latter could be a dust-obscured high-redshift galaxy \citep[e.g.,][]{fudamoto2021} or just spurious.
We further discuss the ALMA sources without NIRCam counterparts in Section~\ref{sec:jwst-dark}.
In Table~\ref{tab:catalog}, we also list the NIRCam source ID (NID) and the observed NIRCam photometry (i.e., without lens correction) in F150W and F444W filters with a $0\farcs32$-diameter aperture taken from \cite{weaver2023}. 

 \startlongtable
\setlength{\tabcolsep}{2pt}
\begin{deluxetable*}{lcccccccccccccc}
\tablecaption{ALMA Source Catalog of the \survey\ survey}
\tablehead{
\colhead{AID} & \colhead{NID} & \colhead{MID} & \colhead{R.A.} & \colhead{Dec}   & \colhead{SNR} & \colhead{Map} & \colhead{PB} & \colhead{$S_{\rm 1.2mm}$} & \colhead{F150W} & \colhead{F444W} & \colhead{$z_{\rm phot}$} &  \colhead{$z_{\rm spec}$} & \colhead{$\mu$} & \colhead{ref.}\\
   & &  & \colhead{deg}   & \colhead{deg}  &  \colhead{}    &  \colhead{}  &  \colhead{}    & \colhead{$\mu$Jy} &  \colhead{mag}     &  \colhead{mag} &  & & & \\
\colhead{(1)} & \colhead{(2)} & \colhead{(3)} &\multicolumn{2}{c}{(4)} & \colhead{(5)} & \colhead{(6)} & \colhead{(7)} & \colhead{(8)} & \colhead{(9)} & \colhead{(10)} & \colhead{(11)}& \colhead{(12)}& \colhead{(13)}& \colhead{(14)}} 
\startdata
\multicolumn{15}{c}{Identified from {\it Wide} maps ($N=65$)} \\     \hline
1 & 3843 & \nodata & 3.6036340084 &$ -30.429562044 $& 5.26 & nat & 0.63 & 274 $\pm$ 52 & 22.37 & 20.39 &  $ 1.53^{+0.09}_{-0.09}$ & \nodata & 1.6 & \nodata \\
2 & 5107 & \nodata & 3.5755829201 &$ -30.424377998 $& 41.37 & nat & 0.58 & 3019 $\pm$ 121 & 28.04 & 21.99 &  $ 4.05^{+0.19}_{-0.20}$ & \nodata & 1.7 & \nodata \\
3 & 5000 & 3928 & 3.5863514538 &$ -30.425044018 $& 15.05 & tap & 0.84 & 797 $\pm$ 53 & 27.23 & 23.68 &  $ 2.97^{+1.20}_{-0.25}$ & $2.99$ & 1.8 & P23 \\
4 & 7360 & 6291 & 3.6056154652 &$ -30.418060598 $& 5.26 & tap & 0.84 & 277 $\pm$ 53 & 21.24 & 20.98 &  $ 2.90^{+0.02}_{-0.03}$ & $2.68$ & 2.3 & P23 \\
5 & 9018 & \nodata & 3.5760482064 &$ -30.413200341 $& 45.16 & tap & 0.84 & 3041 $\pm$ 119 & 27.28 & 23.46 &  $ 2.56^{+0.08}_{-0.08}$ & $2.582$ & 2.0 & K23 \\
6 & 8700 & \nodata & 3.6085747745 &$ -30.414590756 $& 10.48 & nat & 0.85 & 547 $\pm$ 52 & 22.12 & 20.06 &  $ 2.67^{+0.08}_{-0.08}$ & \nodata & 2.2 & \nodata \\
7 & 9103 & \nodata & 3.5650518436 &$ -30.412200215 $& 6.12 & nat & 0.83 & 275 $\pm$ 53 & 23.21 & 20.95 &  $ 1.28^{+0.10}_{-0.05}$ & \nodata & 1.5 & \nodata \\
8 & 14034 & \nodata & 3.5690145916 &$ -30.402797638 $& 6.32 & nat & 0.84 & 293 $\pm$ 53 & 23.87 & 19.66 &  $ 2.43^{+0.07}_{-0.09}$ & $2.582$ & 2.0 & W15 \\
9 & 16840 & \nodata & 3.5723726430 &$ -30.395953862 $& 7.47 & tap & 0.84 & 395 $\pm$ 53 & 24.15 & 21.84 &  $ 3.61^{+0.09}_{-0.15}$ & \nodata & 3.2 & \nodata \\
10 & 17539 & \nodata & 3.6276995116 &$ -30.394278417 $& 21.53 & nat & 0.76 & 1055 $\pm$ 80 & 27.17 & 22.95 &  $ 3.52^{+0.14}_{-0.17}$ & \nodata & 1.4 & \nodata \\
11 & 16790 & \nodata & 3.6005232329 &$ -30.396157886 $& 5.17 & nat & 0.97 & 175 $\pm$ 34 & 22.02 & 20.03 &  $ 0.84^{+0.05}_{-0.05}$ & $0.943$ & 2.0 & W15 \\
12 & 17269 & \nodata & 3.6175368981 &$ -30.395012711 $& 5.98 & nat & 0.76 & 280 $\pm$ 58 & 24.43 & 23.37 &  $ 4.66^{+0.07}_{-0.08}$ & \nodata & 1.6 & \nodata \\
13 & 17477 & 16609 & 3.5667961215 &$ -30.394887615 $& 15.27 & nat & 0.85 & 700 $\pm$ 52 & 25.79 & 21.29 &  $ 2.82^{+0.09}_{-0.11}$ & $3.06$ & 2.3 & P23 \\
14 & 17833 & 16987 & 3.6326634712 &$ -30.393641545 $& 14.00 & nat & 0.75 & 607 $\pm$ 59 & 25.54 & 23.40 &  $ 3.87^{+0.20}_{-0.95}$ & $3.97$ & 1.4 & P23 \\
15 & 18358 & 17516 & 3.6214797562 &$ -30.393109357 $& 13.17 & nat & 0.76 & 563 $\pm$ 58 & 24.97 & 22.44 &  $ 3.63^{+0.12}_{-0.36}$ & $3.22$ & 1.5 & P23 \\
16 & 19562 & 18708 & 3.5809797340 &$ -30.390749405 $& 5.47 & tap & 0.85 & 286 $\pm$ 52 & 19.42 & 19.35 &  $ 0.29^{+0.02}_{-0.04}$ & $0.29$ & 1.0 & P23 \\
17 & 21370 & \nodata & 3.5474110725 &$ -30.388288554 $& 55.22 & nat & 0.78 & 2815 $\pm$ 74 & 24.60 & 20.48 &  $ 2.82^{+0.09}_{-0.12}$ & \nodata & 1.9 & \nodata \\
18 & 21322 & \nodata & 3.5327225543 &$ -30.387320373 $& 6.56 & tap & 0.78 & 374 $\pm$ 57 & 21.56 & 20.43 &  $ 1.37^{+0.83}_{-0.17}$ & \nodata & 1.6 & \nodata \\
19 & 43148 & 42272 & 3.5719637485 &$ -30.383017638 $& 13.54 & tap & 0.88 & 685 $\pm$ 51 & 22.86 & 19.68 &  $ 1.45^{+0.07}_{-0.12}$ & $1.67$ & 2.7 & P23 \\
20 & \nodata & \nodata & 3.5190587771 &$ -30.386326057 $& 8.90 & nat & 0.22 & 1505 $\pm$ 201 & 26.80 & (outside) &  \nodata & \nodata & 1.5$^{\dagger}$ & \nodata \\
21 & 21972 & 21111 & 3.5825066752 &$ -30.385467633 $& 42.03 & nat & 0.98 & 1491 $\pm$ 60 & 24.73 & 21.63 &  $ 2.91^{+0.06}_{-0.08}$ & $3.06$ & 4.2 & P23 \\
22 & 24143 & \nodata & 3.5850010413 &$ -30.381794100 $& 24.50 & nat & 0.81 & 1713 $\pm$ 123 & 23.41 & 20.40 &  $ 2.95^{+0.12}_{-0.09}$ & $3.058$ & 3.0 & M23 \\
23 & 43086 & \nodata & 3.5732506606 &$ -30.383496723 $& 24.55 & nat & 0.90 & 913 $\pm$ 61 & 23.28 & 20.02 &  $ 1.18^{+0.05}_{-0.03}$ & $1.498$ & 2.6 & M23 \\
24 & 23634 & \nodata & 3.6357674798 &$ -30.382387125 $& 11.26 & nat & 0.76 & 599 $\pm$ 59 & 25.04 & 21.48 &  $ 2.49^{+0.14}_{-0.18}$ & \nodata & 1.3 & \nodata \\
25 & 24731 & \nodata & 3.5126765859 &$ -30.380965658 $& 6.94 & nat & 0.30 & 712 $\pm$ 148 & 25.63 & 21.74 &  $ 2.77^{+0.09}_{-0.14}$ & \nodata & 1.5 & \nodata \\
26 & 24823 & \nodata & 3.5920643028 &$ -30.380487291 $& 11.53 & nat & 0.76 & 530 $\pm$ 58 & 26.17 & 21.45 &  $ 2.34^{+0.12}_{-0.10}$ & $2.644$ & 2.3 & M23 \\
27 & 24852 & 23955 & 3.5812925141 &$ -30.380249578 $& 9.12 & nat & 0.87 & 330 $\pm$ 51 & 27.36 & 22.36 &  $ 3.57^{+0.21}_{-0.16}$ & $3.47$ & 2.9 & P23 \\
28 & 26131 & \nodata & 3.5797072291 &$ -30.378412661 $& 21.51 & nat & 0.89 & 938 $\pm$ 72 & 24.07 & 20.52 &  $ 2.34^{+0.13}_{-0.08}$ & $2.409$ & 2.6 & M23 \\
29 & 27243 & \nodata & 3.5599875905 &$ -30.377803997 $& 5.89 & nat & 0.79 & 289 $\pm$ 57 & (masked) &  24.49  & \nodata & \nodata & 1.0$^{\dagger}$ & \nodata \\
30 & 29349 & \nodata & 3.5579207350 &$ -30.377236095 $& 6.71 & tap & 0.79 & 379 $\pm$ 57 & 21.72 & 20.57 &  $ 2.87^{+0.10}_{-0.06}$ & \nodata & 9.0 & \nodata \\
31 & 27891 & \nodata & 3.5823806198 &$ -30.377169986 $& 5.76 & nat & 0.79 & 298 $\pm$ 56 & 26.55 & 23.87 &  $ 4.75^{+0.11}_{-0.14}$ & \nodata & 2.9 & \nodata \\
32 & 28824 & \nodata & 3.5995953836 &$ -30.374707127 $& 7.37 & nat & 0.76 & 383 $\pm$ 58 & 24.95 & 21.48 &  $ 2.04^{+0.09}_{-0.09}$ & \nodata & 1.7 & \nodata \\
33 & 30188 & \nodata & 3.5103148006 &$ -30.375437642 $& 6.05 & nat & 0.68 & 358 $\pm$ 66 & 23.61 & 21.01 &  $ 0.86^{+0.02}_{-0.03}$ & \nodata & 1.3 & \nodata \\
34 & 31310 & 30414 & 3.5543532540 &$ -30.371954754 $& 13.10 & nat & 0.79 & 644 $\pm$ 57 & 23.92 & 20.33 &  $ 2.49^{+0.22}_{-0.25}$ & $2.34$ & 3.8 & P23 \\
35 & 33393 & \nodata & 3.6172781655 &$ -30.368795269 $& 35.60 & nat & 0.76 & 1970 $\pm$ 99 & 20.98 & 20.15 &  $ 1.13^{+0.00}_{-0.00}$ & \nodata & 1.3 & \nodata \\
36 & 31567 & \nodata & 3.5237288439 &$ -30.371468917 $& 8.72 & nat & 0.78 & 330 $\pm$ 57 & 24.11 & 21.50 &  $ 2.41^{+0.40}_{-0.31}$ & \nodata & 2.0 & \nodata \\
37 & 43092 & 42203 & 3.5638034680 &$ -30.367614211 $& 5.38 & tap & 0.91 & 263 $\pm$ 49 & 21.11 & 19.07 &  $ 1.30^{+0.08}_{-0.06}$ & $1.32$ & 2.1 & P23 \\
38 & 36415 & \nodata & 3.5375675466 &$ -30.365649191 $& 6.95 & nat & 0.78 & 342 $\pm$ 57 & 21.95 & 19.82 &  $ 1.65^{+0.09}_{-0.12}$ & \nodata & 2.9 & \nodata \\
39 & 39567 & \nodata & 3.5311608452 &$ -30.361293935 $& 9.22 & nat & 0.78 & 501 $\pm$ 57 & 23.73 & 20.10 &  $ 1.43^{+0.06}_{-0.06}$ & \nodata & 1.8 & \nodata \\
40 & 36484 & 35516 & 3.6006012798 &$ -30.362713723 $& 15.98 & nat & 0.77 & 750 $\pm$ 58 & 25.52 & 21.57 &  $ 2.60^{+0.16}_{-0.23}$ & $2.5$ & 1.6 & P23 \\
41 & 39317 & \nodata & 3.5391400276 &$ -30.360299747 $& 10.55 & tap & 0.79 & 596 $\pm$ 57 & 24.09 & 21.60 &  $ 1.34^{+0.60}_{-0.09}$ & \nodata & 1.9 & \nodata \\
42 & 39328 & 38369 & 3.5362849803 &$ -30.360378268 $& 50.38 & nat & 0.78 & 2677 $\pm$ 88 & 23.25 & 19.71 &  $ 1.47^{+0.03}_{-0.03}$ & $1.89$ & 2.0 & P23 \\
43 & 38303 & \nodata & 3.5211455073 &$ -30.360680181 $& 11.09 & nat & 0.75 & 546 $\pm$ 60 & 21.73 & 19.36 &  $ 1.34^{+0.08}_{-0.13}$ & \nodata & 1.6 & \nodata \\
44 & 38852 & 38163 & 3.5990300334 &$ -30.359756669 $& 12.18 & tap & 0.79 & 682 $\pm$ 56 & 21.43 & 19.35 &  $ 1.27^{+0.10}_{-0.06}$ & $1.36$ & 1.5 & P23 \\
45 & 40516 & \nodata & 3.5938353534 &$ -30.356617123 $& 24.72 & tap & 0.84 & 1540 $\pm$ 119 & 20.93 & 18.81 &  $ 1.28^{+0.07}_{-0.05}$ & \nodata & 1.5 & \nodata \\
46 & 38707 & \nodata & 3.5237433941 &$ -30.359220841 $& 5.86 & nat & 0.75 & 293 $\pm$ 59 & 23.92 & 21.63 &  $ 2.22^{+0.46}_{-0.08}$ & \nodata & 1.7 & \nodata \\
47 & 40522 & \nodata & 3.5926432857 &$ -30.356140111 $& 8.34 & tap & 0.84 & 444 $\pm$ 53 & 25.63 & 22.85 &  $ 2.59^{+0.14}_{-0.37}$ & \nodata & 1.7 & \nodata \\
48 & 43129 & \nodata & 3.5583154162 &$ -30.354968641 $& 23.83 & nat & 0.83 & 1222 $\pm$ 93 & 26.75 & 24.23 &  $ 3.52^{+0.11}_{-0.32}$ & \nodata & 1.8 & \nodata \\
49 & 41848 & \nodata & 3.5463708998 &$ -30.353475868 $& 6.77 & nat & 0.85 & 277 $\pm$ 52 & 25.36 & 22.03 &  $ 2.30^{+0.25}_{-0.13}$ & \nodata & 1.7 & \nodata \\
50 & 42323 & \nodata & 3.5491493630 &$ -30.352244608 $& 39.28 & nat & 0.83 & 1871 $\pm$ 71 & 24.03 & 20.56 &  $ 2.30^{+0.20}_{-0.04}$ & \nodata & 1.7 & \nodata \\
51 & 44013 & \nodata & 3.5305552202 &$ -30.352316924 $& 7.60 & nat & 0.78 & 317 $\pm$ 57 & 24.21 & 21.88 &  $ 1.20^{+0.14}_{-0.20}$ & \nodata & 1.5 & \nodata \\
52 & 53471 & \nodata & 3.5612865835 &$ -30.326962648 $& 5.36 & nat & 0.81 & 216 $\pm$ 40 & 27.44 & 23.76 &  $ 3.82^{+0.18}_{-0.63}$ & \nodata & 1.4 & \nodata \\
53 & 50074 & \nodata & 3.5555550052 &$ -30.335626961 $& 6.31 & nat & 0.83 & 261 $\pm$ 54 & 24.57 & 23.04 &  $ 3.64^{+0.09}_{-0.11}$ & \nodata & 1.5 & \nodata \\
54 & 46628 & \nodata & 3.5528433792 &$ -30.344172744 $& 8.76 & nat & 0.83 & 441 $\pm$ 54 & 21.68 & 20.12 &  $ 1.56^{+0.12}_{-0.28}$ & \nodata & 1.5 & \nodata \\
55 & 46360 & \nodata & 3.5641530944 &$ -30.344466238 $& 17.08 & nat & 0.83 & 970 $\pm$ 93 & 23.99 & 21.53 &  $ 3.20^{+0.08}_{-0.08}$ & \nodata & 1.6 & \nodata \\
56 & 52675 & \nodata & 3.5611003202 &$ -30.330042986 $& 8.83 & nat & 0.83 & 415 $\pm$ 53 & 22.65 & 19.86 &  $ 1.47^{+0.10}_{-0.11}$ & \nodata & 1.3 & \nodata \\
57 & 22699 & \nodata & 3.5566688470 &$ -30.384298577 $& 4.51 & tap & 0.78 & 255 $\pm$ 57 & 23.30 & 21.54 &  $ 2.51^{+0.09}_{-0.07}$ & \nodata & 2.8 & \nodata \\
58 & 24619 & \nodata & 3.6384449479 &$ -30.381212884 $& 4.59 & tap & 0.67 & 306 $\pm$ 67 & 19.71 & 19.22 &  $ 0.89^{+0.05}_{-0.07}$ & \nodata & 1.2 & \nodata \\
59 & 29373 & \nodata & 3.5582540814 &$ -30.374436946 $& 4.68 & tap & 0.79 & 264 $\pm$ 56 & 22.61 & 20.67 &  $ 2.55^{+0.06}_{-0.05}$ & \nodata & 4.3 & \nodata \\
60 & \nodata & \nodata & 3.5923715243 &$ -30.368396301 $& 4.63 & tap & 0.78 & 265 $\pm$ 57 & $>$ 29.35 &  $>$ 29.41 &  \nodata & \nodata & 2.2$^{\dagger}$ & \nodata \\
61 & 52679 & \nodata & 3.5580404441 &$ -30.331104285 $& 4.91 & tap & 0.83 & 262 $\pm$ 53 & 22.18 & 20.75 &  $ 1.32^{+0.42}_{-0.58}$ & \nodata & 1.3 & \nodata \\
62 & 49492 & 48540 & 3.5456886277 &$ -30.337677783 $& 4.47 & tap & 0.82 & 243 $\pm$ 54 & 23.99 & 21.14 &  $ 1.62^{+0.11}_{-0.12}$ & $1.65$ & 1.4 & P23 \\
63 & 48152 & 47271 & 3.5671705172 &$ -30.341419478 $& 4.41 & tap & 0.83 & 236 $\pm$ 53 & 24.73 & 21.66 &  $ 1.90^{+0.17}_{-0.23}$ & $1.84$ & 1.5 & P23 \\
64 & 52684 & \nodata & 3.5618008005 &$ -30.330958075 $& 5.37 & tap & 0.84 & 286 $\pm$ 53 & 18.77 & 18.72 &  $ 0.65^{+0.04}_{-0.04}$ & $0.644$ & 1.2 & K22 \\
65 & 52426 & \nodata & 3.5685704345 &$ -30.329641642 $& 4.68 & tap & 0.66 & 1137 $\pm$ 242 & 27.29 & 28.10 &  $ 0.65^{+0.08}_{-0.09}$ & \nodata & 1.2 & \nodata \\ \hline
\multicolumn{15}{c}{Additionally identified from {\it Deep} maps ($N=4$)} \\    \hline
66 & 10578 & \nodata & 3.5938124804 &$ -30.408262675 $& 4.93 & nat & 0.89 & 150 $\pm$ 47 & 29.36  &  $>$ 29.47 & $ 9.89^{+1.22}_{-9.20}$     &  \nodata          & 3.4  & \nodata \\
67 & 11129 & \nodata & 3.6017503242 &$ -30.407851038 $& 5.20 & nat & 0.90 & 157 $\pm$ 45 & 22.48 & 20.45 &  $ 1.08^{+0.04}_{-0.05}$ & $1.1$ & 2.9 & K22 \\
68 & 23405 & \nodata & 3.5932584449 &$ -30.384374741 $& 5.01 & nat & 0.82 & 166 $\pm$ 51 & 18.76 & 18.58 &  $ 0.31^{+0.03}_{-0.03}$ & $0.296$ & 1.0 & K22 \\
69 & 15750 & 14267 & 3.5902747340 &$ -30.400439262 $& 4.21 & tap & 0.85 & 175 $\pm$ 44 & 18.45 & 18.15 &  $ 0.56^{+0.05}_{-0.06}$ & $0.50$ & 3.4 & P23
\enddata
\tablecomments{
(1): ALMA Source ID.
(2): NIRCam Source ID in the DR2 catalog \citep{weaver2023} of the NIRCam counterparts identified with the $1\farcs0$ search radius. 
(3): MSA ID for the sources observed with NIRSpec prism (S.~Price in prep.). 
(4): Source coordinate of the ALMA continuum peak in the natural map.
(5): Signal-to-noise ratio (SNR) of the peak pixel in the natural or the tapered map (the map showing the higher SNR being adopted).  
(6): Map used for the source identification, where the source SNR is maximized. ``nat'' and ``tap'' denote the natural-weighted and $uv$-tapered maps, respectively.  
(7): Relative sensitivity response to the deepest of the mosaic. 
(8): Source flux density at 1.2 mm without the lens correction, measured either by the peak pixel in the natural map, the peak pixel in the tapered map, a $1\farcs0$-radius aperture in the natural map, or the 2D elliptical Gaussian fitting with CASA {\sc imfit} (the measurement showing the highest SNR being adopted). 
(9--10): Aperture magnitude with a $0\farcs32$ diameter in the NIRCam/F150W and F444W filters without the lens correction, measured in \cite{weaver2023}. For non-detection, we place the 2$\sigma$ upper limit from the error estimate for the closest nearby source in the catalog.  
(11): Photometric redshift using \eazy\ and \prospector\ estimated in B.~Wang et al. in (prep.). 
(12): Spectroscopic redshift presented in literature (K22; \citealt{kokorev2022}, K23; \citealt{kokorev2023}, L17; \citealt{laporte2017}, M23; \citealt{arancibia2023}, \citealt{wang2015}) and the latest NIRSpec spectroscopy (P23; S.~Price et al. in preparation).  
(12): Magnification factor based on the mass model presented in \cite{furtak2023a}.  
}
$\dagger$ We assume $z=2.0$ and $z=9.0$ for ID20 and ID60, respectively (see text). For ID29, we interpret it as a cluster member galaxy based on its location, color, and morphology.  
\label{tab:catalog}
\end{deluxetable*}

We note that the misidentification of the NIRCam counterpart may also occur due to chance projection.
Based on the NIRCam source catalog from \cite{weaver2023}, the number density of the NIRCam sources is estimated to be $\sim$0.26arcsec$^{-2}$ down to F150W of 29.0mag, equal to the faintest NIRCam counterpart among our ALMA sources.
This surface density yields a probability of chance projection \citep{downes1986} being $\sim$0--30\% with the spatial offsets of $0\farcs0-0\farcs7$. By integrating these probabilities, we obtain the expected number of chance projections to be $\sim2.3$ among our ALMA sources, which is comparable to the number of negative sources above our SNR threshold (see Figure~\ref{fig:pn_hist}). Though this is the regime of small-number statistics, this suggests that one or two spurious sources might be included in our ALMA sources, especially in those with large spatial offsets from their NIRCam counterparts.

\begin{figure*}[t!]
\includegraphics[trim=0cm 0cm 0cm 0cm, clip, angle=0,width=1.\textwidth]{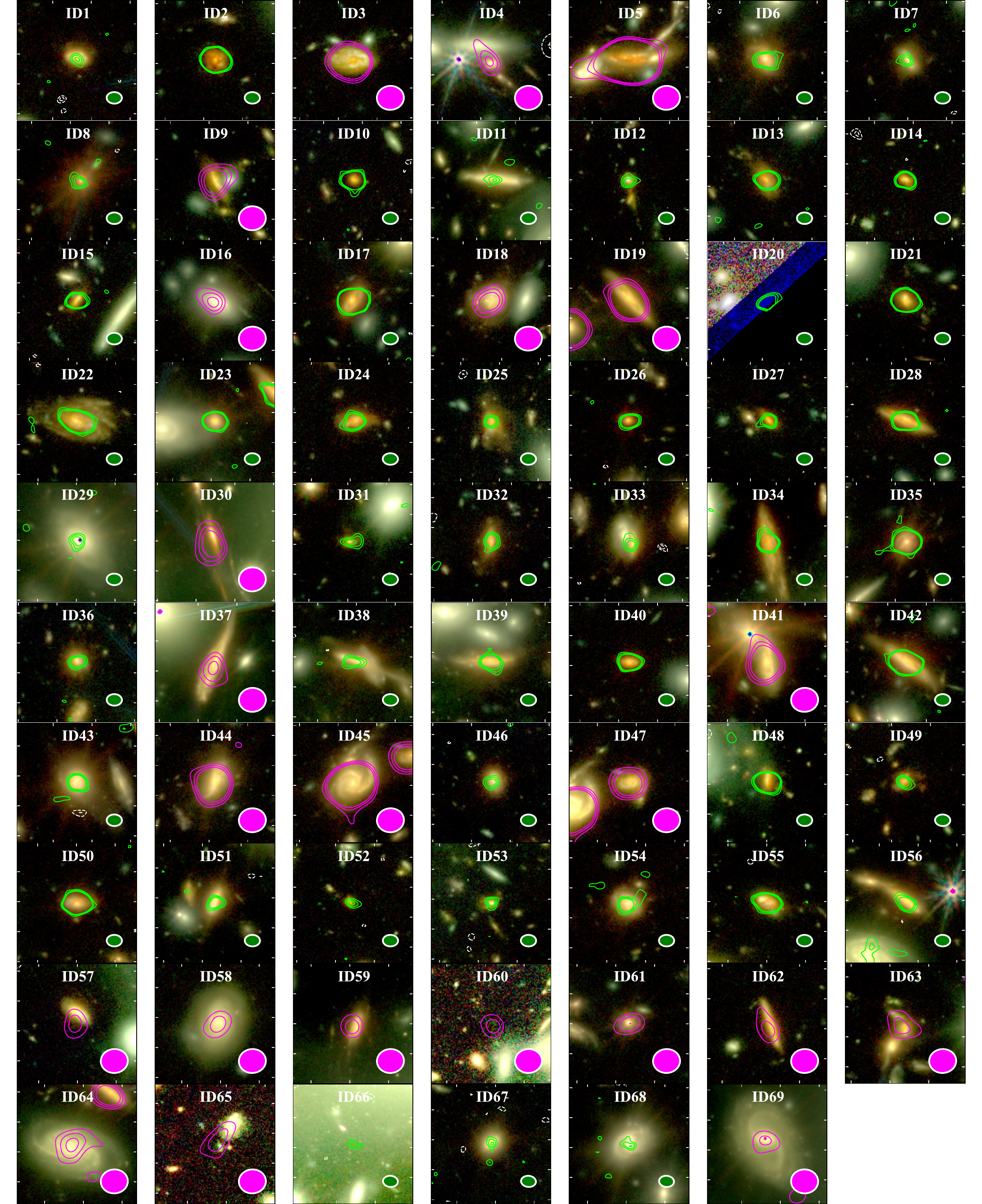}
\vspace{-0.4cm}
\caption{
RGB NIRCam color cutouts ($8"\times8"$) for the 69 ALMA sources identified in the survey (R: F444W, G: F277W, B: F150W). The green and magenta contours represent the $3\sigma$, $4\sigma$, and $5\sigma$ significance levels of the ALMA 1.2-mm continuum in the natural and tapered maps, respectively, with the map showing the higher SNR being adopted.
The white contours denote the $-3\sigma$, $-4\sigma$, and $-5\sigma$ significance levels.
The ellipse displayed at the bottom right represents the ALMA synthesized beam.
\label{fig:postage}}
\end{figure*}

\begin{figure*}[t]
\includegraphics[trim=0cm 0cm 0cm 0cm, clip, angle=0,width=1.\textwidth]{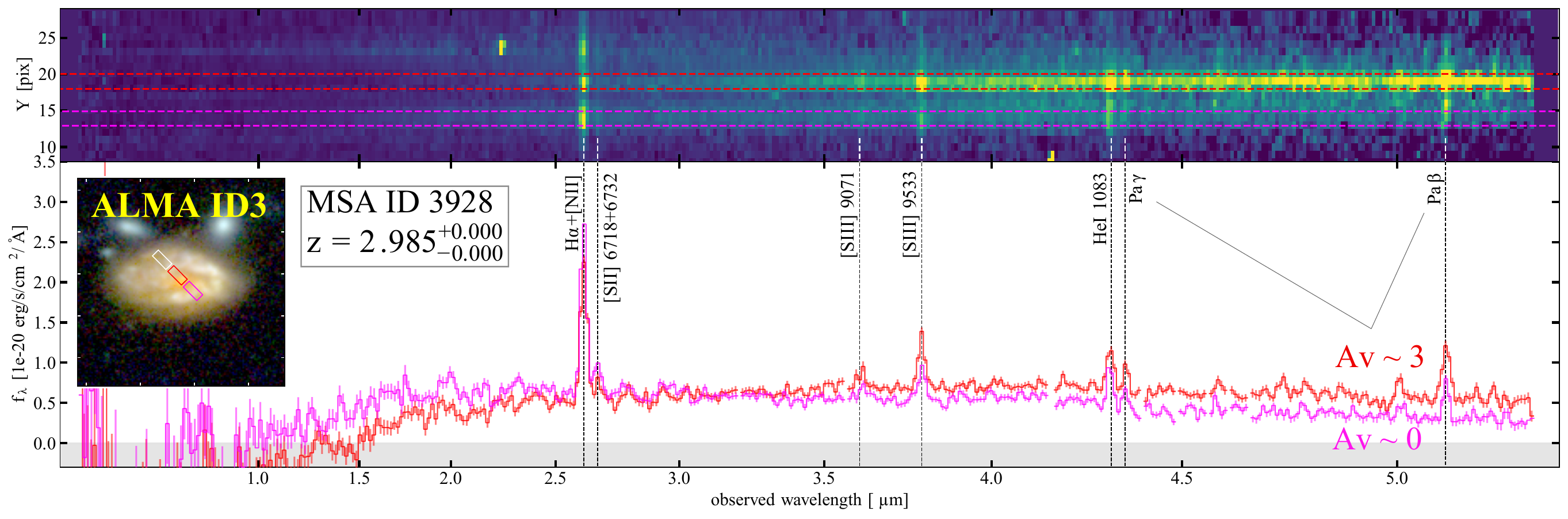}
\vspace{-0.4cm}
 \caption{
Example of the 2D (\textit{top}) and 1D (\textit{bottom}) \jwst/NIRSpec prism spectra taken for the ALMA sources in \survey. 
The inset panel shows the NIRCam RBG $5''\times5''$ cutout, where the three shutter positions are overlaid. 
The inset label shows the MSA ID and our best-fit redshift estimate from the prism spectrum. 
To avoid self-subtraction, we use the global background subtraction for the 2D and 1D spectra instead of the standard three-shutter nod method for this source in this paper (see text). 
The red and magenta curves (horizontal dashed lines) in the bottom (top) panel indicate the 1D spectra (extraction apertures) for the data taken from the shutters highlighted with the red and magenta rectangles in the inset panel, respectively. 
The black vertical lines denote wavelengths of the emission lines detected at SNR $\geq$ 4.0. 
The Balmer decrement via Pa$\gamma$/Pa$\beta$ shows the spatial variation of the dust attenuation, demonstrating the power of NIRSpec MSA also in the spatially-resolved fashions. 
\label{fig:nirspec_id3}}
\end{figure*}

\subsection{NIRSpec prsim spectroscopy}
\label{sec:nirspec}

Following the survey design of UNCOVER \citep{bezanson2022}, a systematic NIRSpec follow-up was performed for NIRCam-selected galaxies with the Multiobject Spectroscopy (MOS) mode in July-August 2023. Based on exciting high-redshift source candidates (e.g., very high-redshift galaxies, faint AGNs, quiescent galaxies, strongly magnified and multiply imaged sources), the MOS configurations with the multi-shutter array (MSA) were designed to maximize the number of the observed exciting candidates. 
Among the NIRCam counterparts of the ALMA sources, 
seventeen sources were assigned to the MSA masks. 
Multiple emission lines have been detected from all these seventeen ALMA sources, securely determining their spectroscopic redshift ($z_{\rm spec}$). 
The full description of the NIRSpec data reduction and the redshift estimates will be presented in S.~Price et al. in preparation (see also \citealt{goulding2023, bwang2023, furtak2023c, atek2023b, fujimoto2023c, kokorev2023b, burgasser2023, greene2023}). 
In Table~\ref{tab:catalog}, we also list MSA ID (MID) for the seventeen ALMA sources.

As an example of the redshift estimate with the NIRSpec prism, Figure~\ref{fig:nirspec_id3} shows the 1D and 2D prism spectrum taken from one of the ALMA sources (AID3). 
The inset panel presents the MSA shutter positions overlaid on the NIRCam RGB image. 
Because the source morphology in NIRCam indicates that the emission from the source can be included in all three shutters of the MSA, here we perform a global background subtraction by using a nearby empty shutter, instead of the standard three shutter-nod method, and extract the 1D spectra from the central (red curve) and outer disk regions (magenta curve).
From a Spectral Energy Distribution (SED) template fitting using \texttt{EAZY} \citep{brammer2008} in the same manner as \cite{fujimoto2023c}, the source redshift is determined at $z_{\rm spec}=2.985$. 
Fixing the redshift, a subsequent spline fitting to the continuum combined with the single Gaussian for each emission line yields the successful multiple line detection at SNR $\geq4$, including H$\alpha$+\nii, \sii, \siii, HeI, Pa$\gamma$, and Pa$\beta$ in both regions. The dust attenuation is estimated via the Balmer decrement of Pa$\gamma$/Pa$\beta$ with the \cite{calzetti2000} law, resulting in $A_{\rm V}\sim3$ and $A_{\rm V}\sim0$ in the central and outer disk region, respectively. 
This provides the spectroscopic evidence of the physical association between the ALMA emission and the significant dust attenuation in a high-redshift galaxy and also unveils its central concentration, possibly related to the process of the massive bulge formation. 
Note that AID3 is detected with a better SNR in the tapered map (Table~\ref{tab:catalog}), suggesting the presence of extended dust continuum. Meanwhile, the NIRSpec results indicate less dust obscuration in the outer regions. As shown in the inset panel of Figure~\ref{fig:nirspec_id3}, the NIRCam RGB image of AID3 reveals a complex morphology, consisting of dusty red regions and less-obscured blue regions, where the other MSA shutter coincides with the blue regions. This likely reflects the less obscuration measured in the outer MSA shutter, consistent with this geometric interpretation.
The joint ALMA and NIRSpec MSA analysis demonstrates the power of determining the source redshift and gaining insights into dust-obscured properties in high-redshift galaxies, also in a spatially resolved manner.  
Further NIRSpec analyses and results for dusty galaxies in A2744 will be presented in separate papers (S.~Price et al. in preparation; C.~Williams et al. in preparation).

\subsection{Physical properties}
\label{sec:phy_properties}

To obtain the physical properties constrained owing to the rich \jwst\ and \hst\ photometry available in A2744, we further cross-match the 67 NIRCam counterparts of our ALMA sources with the physical parameter catalog constrained with all available \hst\ and \jwst\ data in A2744 and the ALMA photometry obtained from the \survey\ survey \citep{bwang2024}. The details of the SED analysis is presented in \cite{bwang2024}, and here we briefly explain the SED model procedures. 

The main products of the catalog are inferred using the \prospector\ Bayesian inference framework \citep{johnson2021}, leveraging simple stellar populations (SSPs) from FSPS \citep{conroy2010} with a Chabrier initial mass function (IMF; \citealt{Chabrier2003}), where MIST isochrones \citep{choi2016,dotter2016} and the MILES \citep{sanchez2006} stellar library are adopted. The complex stellar populations are modeled using \prospector-$\beta$ \citep{wang2023}, 
with a star-formation history (SFH) modeled as mass formed in 7 logarithmically-spaced time bins. 
Dust is described using a two-component model \citep{charlot2000} with a flexible dust attenuation curve \citep{noll2009}, and the dust emission \citep{draine2007}, stellar metallicity, and mid-infrared AGNs are all included in the fit. 
\prospector-$\beta$ uses informed priors for redshift and mass based on the observed evolution of the galaxy stellar mass function, and star formation history based on the observational result that massive galaxies form earlier while low-mass galaxies form later, to fully exploit the \prospector\ Bayesian framework.
Simultaneously, the lensing magnification is accounted for during the parameter fitting process because the scale-dependent priors necessitate a self-consistent treatment of the magnification factor, $\mu$. A sampling of the posterior space is performed using the nested sampler \texttt{dynesty}, and an SPS model emulator, \texttt{parrot}, is employed to decrease runtime \citep{mathews2023}. 

An independent SED fitting using \eazy\ \citep{brammer2008} has also been performed. 
\eazy\ is a flexible galaxy photometric redshift code that fits observed SEDs using a non-negative linear combination of templates by minimizing the $\chi^2$ statistics. To avoid the inclusion of nonphysical contributions from templates older than the universe at a given redshift, the updated \texttt{sfhz} template set is adopted. This set includes a template from a $z\sim8$ extreme emission-line galaxy \citep{carnall2023}, enabling the model to account for high-redshift observations.
Following standard \eazy\ methodology, the photometric zero-point offsets \citep{muzzin2013,skelton2014,whitaker2014,straatman2016} and common priors on the magnitude and the UV slope to regulate both low- and high-$z$ solutions are applied to the fitting.

In Table~\ref{tab:catalog}, we list our photometric redshift estimates $z_{\rm phot}$ mainly from \prospector. 
In a few cases, the two $z_{\rm phot}$ estimates deviate by more than 1.0, and in such cases, we adopt the $z_{\rm phot}$ estimate from \eazy, which is expected to provide stable fits due to its smaller number of parameters. 
We also cross-match the NIRCam counterparts with the spectroscopic redshift ($z_{\rm spec}$) compilation in A2744 from the literature \citep{fujimoto2023b} and the latest NIRSpec results (S.~Price in preparation) and list $z_{\rm spec}$ values when matched. 
This yields 27 sources whose redshifts have been spectroscopically determined. 

\begin{figure}
\begin{center}
\includegraphics[trim=0cm 0cm 0cm 0cm, clip, angle=0,width=0.5\textwidth]{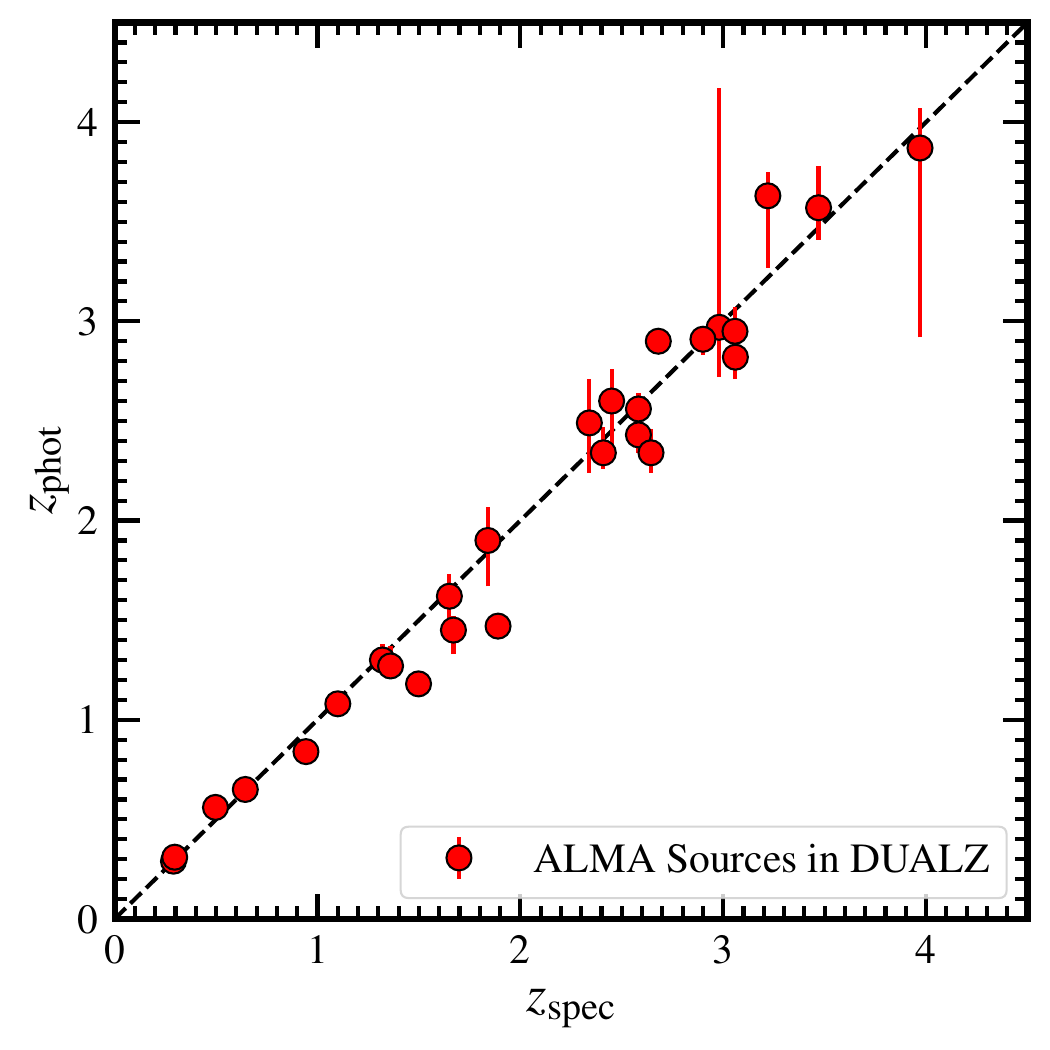}
\caption{
Comparison between $z_{\rm phot}$ and $z_{\rm spec}$ for the spec-$z$ confirmed 27 ALMA sources identified in the \survey\ survey. All 27 sources fall in $|z_{\rm spec}-z_{\rm phot}|/(1+z_{\rm spec}) < 0.15$. 
\label{fig:zspec-zphot}}
\end{center}
\end{figure}

In Figure~\ref{fig:zspec-zphot}, we compare $z_{\rm phot}$ and $z_{\rm spec}$ for the 27 spec-$z$ confirmed ALMA sources identified in the \survey\ survey. 
We find that all the $z_{\rm phot}$ estimates show excellent agreement with $z_{\rm spec}$: 
all 27 sources fall in the range of $|z_{\rm spec}-z_{\rm phot}|/(1+z_{\rm spec}) < 0.15$, indicating the outlier fraction described in \cite{hildebrandt2012} to be zero.  
This validates our SED fitting methods, likely owing to the comprehensive data sets of \hst, \jwst, and ALMA. 
In the following analysis, we use $z_{\rm spec}$ if available, and otherwise, we use $z_{\rm phot}$.
For AID20 that slightly falls outside the NIRCam footprint, we assume $z=2.0$ based on the median redshift estimated to be $z\sim2$ in previous ALMA studies for faint mm sources \citep[e.g.,][]{dunlop2017, yamaguchi2020, aravena2020, gomez2021, fujimoto2023b}. 
For AID60 that lacks a NIRCam counterpart, we assume $z=9.0$, because we find that $z\gtrsim9$ is required to explain the uniquely NIR-faint and mm-detected color property by scaling and shifting the best-fit SED template of the NIR-faint dusty galaxy, REBELS-29-2, at $z=6.68$ \citep{fudamoto2021}. We further discuss AID60 in Section~\ref{sec:jwst-dark}, including the possibility of it being spurious.  
Based on the above procedure of the redshift estimate and the latest mass model in A2744 presented in \cite{furtak2023a}, we also summarize the magnification $\mu$ estimate for the 69 ALMA sources in Table~\ref{tab:catalog}. 

By following these procedures, we estimate the redshift, magnification, and intrinsic 1.2-mm flux after the lens correction ($S_{\rm 1.2mm}^{\rm int}$) for our ALMA sources to be $z=$0.29--9.89, $\mu=$1.0--9.3, and $S_{\rm 1.2mm}^{\rm int}=$0.04--1.65~mJy. 
The intrinsically faintest source is AID30 ($z_{\rm phot}=2.87^{+0.10}_{-0.06}$, $\mu=9.0$, $S_{\rm 1.2mm}^{\rm int}$=0.04~mJy), and we confirm its strongly distorted morphology in the NIRCam maps (see Figure~\ref{fig:postage}).  
The typical properties of our ALMA sources are characterized by these median values of $z=2.30$, $\mu=1.7$, and $S_{\rm 1.2mm}^{\rm int}=0.24$~mJy, respectively.
The full SED results (e.g., SFR, $M_{\rm star}$) is presented in \cite{bwang2024}.

\section{Initial Results}
\label{sec:result}

\subsection{Morphology of the ALMA sources}
\label{sec:morphology}

The median flux density at 1.2mm for the 69 ALMA sources, after lens correction, is estimated at 0.24~mJy (Section~\ref{sec:phy_properties}). 
While our sample also includes some of the intrinsically bright sources, this indicates that the ALMA sources in \survey\ are typically at least $\sim5$ times fainter than the classical dusty starburst populations identified in submm/mm single-dish surveys, so-called SMGs ($S_{\rm 1mm}\gtrsim1$~mJy). 
The great sensitivity and spatial resolution of the NIRCam maps, homogeneous at $\sim1$--5~$\mu$m in UNCOVER, offer a unique opportunity to examine the NIR morphology for these faint ALMA mm sources and thus lead to the \jwst\ insight into what triggers the dusty star-forming activities in these faint mm sources, and a potential difference from that of SMGs.

From visual inspection in Figure~\ref{fig:postage}, we find that almost all NIRCam counterparts of the ALMA sources show undisturbed morphology, denoted either by disk or spheroid. Although several sources  (AID12, AID33, AID38, AID55, AID56, AID63) have another galaxy nearby, which may indicate merging galaxies, the dust-emitting region traced by the ALMA contours always arises from the central red-colored NIRCam galaxy with the undisturbed morphology, instead of the collisional plane with the potential merging galaxies. 
This leads to the conclusion that the dominant populations for the faint ALMA mm sources are represented by undisturbed disk and spheroid galaxies, with the merger fraction among them being $<$10\% (=6/70) at maximum.
Such undisturbed, smooth stellar structures have also been reported in recent MIRI observations for faint ALMA mm sources in HUDF \citep{boogaard2023} and NIRCam observations for lensed ALMA sources \citep{cheng2023}. 
The low merger fraction among the faint ALMA mm sources aligns with recent ALMA and \hst\ studies \citep[e.g.,][]{fujimoto2018}, indicating that the origin of the faint mm emission is not the merging event, but less violent mechanisms such as gas-rich disk instability \citep[e.g.,][]{fujimoto2018, rujopakarn2019}.
In contrast, around 80\% of the SMGs at $z=1-3$ have disturbed morphologies in the rest-frame optical wavelength with \hst/F160W \citep{chen2015}. A high merger fraction is also reported among brighter mm sources ($S_{\rm 1.1mm}\sim1$~mJy) identified in GOODS-S \citep{franco2020}. These results may suggest a critical transition in the submm/mm flux density, above which the merging event becomes the dominant mechanism to yield such bright fluxes in the subm/mm wavelengths. 
We note, however, that \cite{gillman2024} find little difference in the NIRCam and MIRI morphologies of 80 SMGs compared to control samples of field galaxies with matched specific SFR and $M_{\rm star}$. The authors conclude that the triggering mechanisms of SMGs are driven by a combination of their high gas fractions and gravitational instabilities. This result indicates that dust obscuration may still be significant even in the rest-frame optical, potentially leading to an overestimation of the merger fraction in previous HST-based studies that identified disturbed morphologies.

Noteworthy is that a certain number of the sources (AID8, AID10, AID23, AID26, AID25, AID35, AID36, AID40, AID43, AID61) shows a very bright, point-like component in F444W at the center of the galaxy. Albeit the visual inspection, this might indicate the emergence of an active galactic nucleus (AGN), which is routinely identified in recent \hst\ and \jwst\ studies, despite their small survey volumes \citep[e.g.,][]{morishita2020,fujimoto2022, onoue2022, kocevski2023, furtak2022b, endsley2023, furtak2023b, harikane2023c, labbe2023, matthee2023}, or a very compact bulge  \citep[e.g.,][]{lelli2021, tchen2022, rujopakarn2023, boogaard2023}. The dedicated SED analysis will be presented in a separate paper.

\subsection{Optical--NIR dark galaxies}

\subsubsection{\hst-dark galaxies}
\label{sec:hst-dark}

\begin{figure}
\begin{center}
\includegraphics[trim=0cm 0cm 0cm 0cm, clip, angle=0,width=0.5\textwidth]{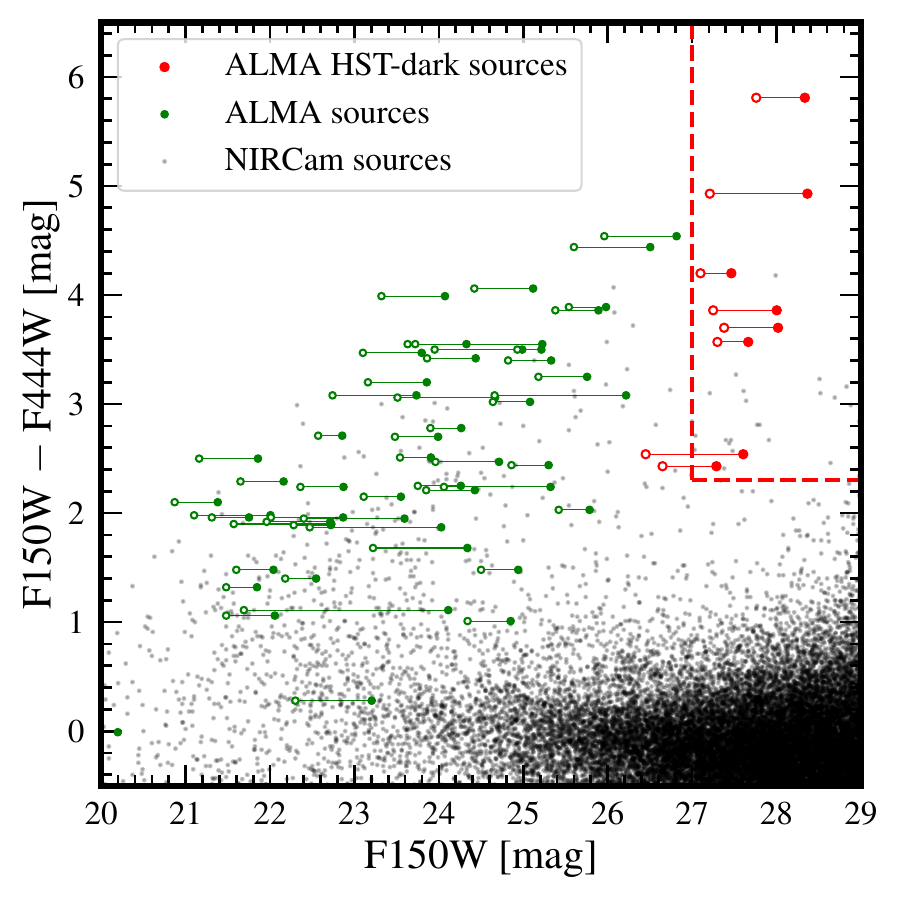}
\caption{
Color--magnitude diagram (F150W vs F150W-F444W).
Filled and open green circles denote our ALMA sources before and after lens correction, respectively.
Filled and open red circles indicate the eight ALMA sources whose colors meet the selection criteria (red dashed line; F150W$>$27~mag, F150W-F444W$>$2.3~mag) of the \hst-dark galaxies, similar to the method of \cite{barrufet2022}.
Black circles represent the NIRCam sources detected in UNCOVER \citep{weaver2023}.
\label{fig:hst-dark_color}}
\end{center}
\end{figure}

\begin{figure*}
\includegraphics[trim=0cm 0cm 0cm 0cm, clip, angle=0,width=1.\textwidth]{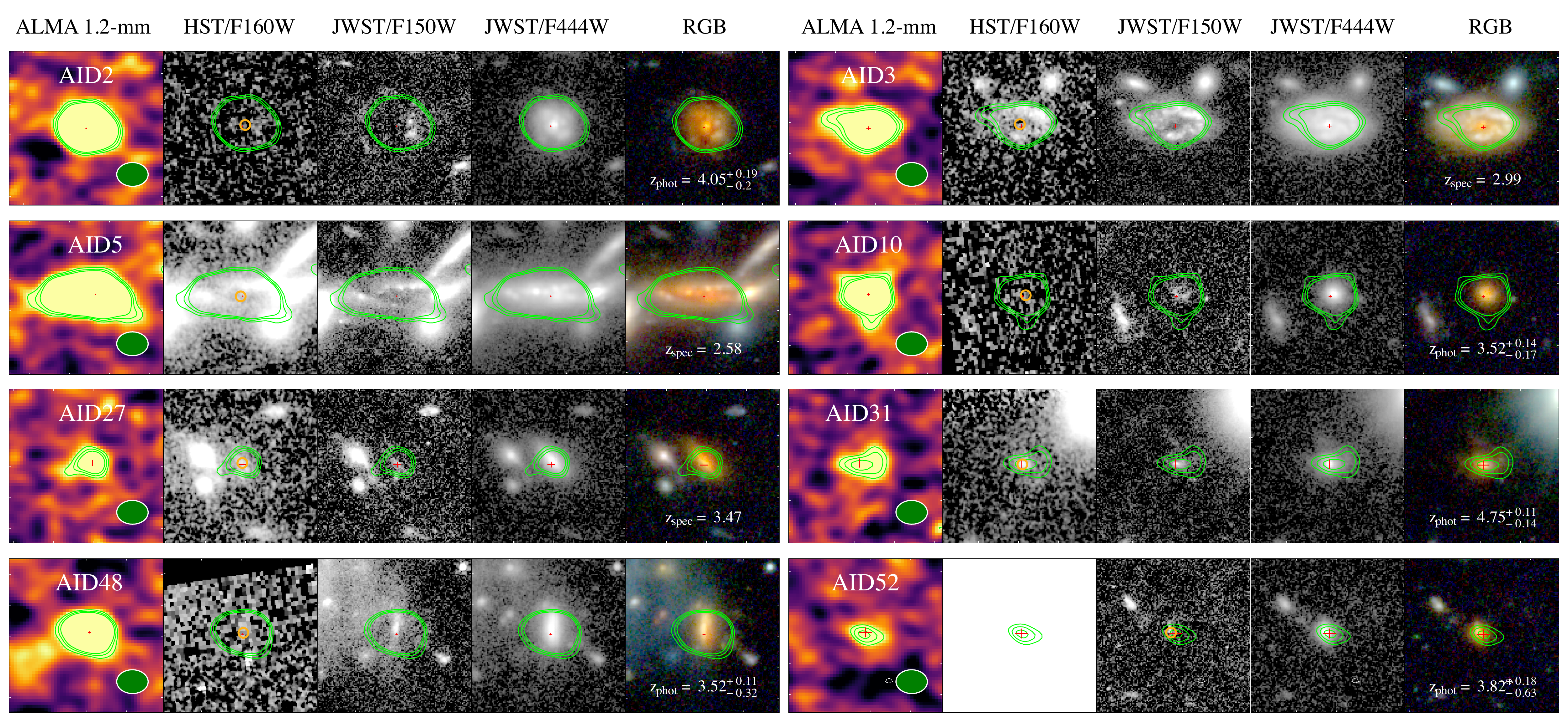}
\vspace{-0.4cm}
\caption{
RGB NIRCam color cutouts ($5''\times5''$) for the eight \hst-dark ALMA sources, selected with the color-magnitude criteria (Figure~\ref{fig:hst-dark_color}).  
For the selection, we use the $0\farcs32$-diameter aperture photometry in NIRCam to fairly investigate the NIR color properties in the compact dust-emitting regions, and thus, the outer regions are clearly visible in \hst/F160W in some cases. 
Using the intrinsic magnitude after the magnification correction may also make some sources clearly visible in \hst.  
The RGB filter assignment is identical to Figure~\ref{fig:postage}.
Green contours represent the $3\sigma$, $4\sigma$, and $5\sigma$ significance levels of the ALMA 1.2-mm continuum in the natural map.
The red cross and its bar scale indicate the ALMA peak pixel position and its positional uncertainty, calculated based on the ALMA beam size and SNR (see text).
The ellipse displayed in the bottom panel represents the ALMA synthesized beam. 
The brown circle denotes the $0\farcs32$-diameter aperture used for the NIRCam photometry \citep{weaver2023}.
\label{fig:hst-dark}}
\end{figure*}

Recent ALMA studies have reported submm/mm sources that are very faint in optical--NIR bands ($\gtrsim26-27$mag) and sometimes invisible even in deep {\it HST}/F160W images, known as {\it HST}-dark galaxies \citep[e.g.,][]{simpson2014, fujimoto2016, franco2018, yamaguchi2019, twang2019, williams2019, casey2019, romano2020, umehata2020, fudamoto2021, gomez2021, talia2021, sun2021, smail2021, fujimoto2022, xiao2022, shu2022, manning2022, guilietti2023}. These \hst-dark galaxies are likely to be heavily dust-obscured ($A_{\rm V}\gtrsim2-5$) massive galaxies at $z\gtrsim3$ \citep[e.g.,][]{twang2019}, and some have been spectroscopically confirmed even at $z>7$ \citep{fudamoto2021}, underlining the incompleteness of galaxy surveys at $z\gtrsim3$ based on imaging at $\lesssim2$$\mu$m \citep[e.g.,][]{fujimoto2023b}. 
With the great sensitivity of NIRCam at $\sim1$--5~$\mu$m, recent \jwst\ studies have overcome these galaxies' inherently low luminosities and further characterized the nature of these {\it HST}-dark galaxies \citep[e.g.,][]{kokorev2023, barrufet2023a, mckinney2023b, smail2023}.

Benefiting from the first homogeneous wide ALMA and \jwst\ maps available in \survey\ and the support of lensing, we also search for {\it HST}-dark galaxies among our 69 ALMA sources. Slightly modifying the classical color cut of F160W$-[4.5]>2.3$ \citep[e.g.,][]{caputi2012, wang2016}, \cite{barrufet2023a} defined the \hst-dark galaxies using the criteria of F150W$-$F444W$>2.3$~mag and F160W$>27$~mag. Similarly, we regard the sources with F150W$-$F444W$>$2.3~mag and F150W$>$27~mag (after lens correction) as \hst-dark galaxies in this paper. Since the dust-emitting regions in high-$z$ galaxies have been measured to be compact (FWHM$\sim0\farcs2-0\farcs3$; e.g., \citealt{ikarashi2015, simpson2015a, hodge2016, fujimoto2017, fujimoto2018, gullberg2019}), we use the NIRCam photometry with a $0\farcs32$-diameter measured in \cite{weaver2023} to fairly investigate the same regions.

Figure~\ref{fig:hst-dark_color} shows the F150W$-$F444W vs F150W distribution of our 69 ALMA sources and the NIRCam-detected sources in UNCOVER. We identify eight ALMA sources (AID2, AID3, AID5, AID10, AID27, AID31, AID48, and AID52) that satisfy the above color criteria for the \hst-dark galaxies. The redshift and $M_{\rm star}$ for these galaxies are estimated to be $z=2.58-4.75$ and $\log(M_{\rm star}/M_{\odot})=$9.81--10.66, with medians of $z=3.52$ and $\log(M_{\rm star}/M_{\odot})=9.96$, based on the \prospector\ fit (Section~\ref{sec:phy_properties}). Compared to the median values of $z=3.80$ and $\log(M_{\rm star}/M_{\odot})=10.60$ estimated in the \hst-dark galaxies identified in the field survey presented in \cite{twang2019}, our \hst-dark ALMA galaxies are less massive, likely due to the deep detection limits enabled by gravitational lensing support.
This suggests that the \hst-dark galaxy population exists not only at the massive end ($\log(M_{\rm star}/M_{\odot})\gtrsim10.5$; e.g., \citealt{weaver2022}) of the stellar mass function at these epochs,
but also at the less-massive regime down to $\log(M_{\rm star}/M_{\odot})=9.8$ at least, indicative of an additional component contributing to the cosmic SFR density at $z\gtrsim3$ across a wide mass range \citep[see also e.g.,][]{xiao2022, algera2023}.

In Figure~\ref{fig:hst-dark}, we also display ALMA, \hst, and \jwst\ cutouts for these eight \hst-dark ALMA sources, where the ALMA source position is marked by the red cross.
We find that four sources (AID3, AID5, AID27, AID48) likely show edge-on disk morphology, while the other sources (AID2, AID10, AID31, AID52) appear to exhibit face-on disk morphology.
This suggests that the heavy dust obscuration among the \hst-dark ALMA galaxies is not always caused by their inclination, but may also be associated with compact dusty star-forming regions \citep{smail2021, lorenz2023, gomez2023}.
However, it is worth mentioning that dust lanes are clearly observed in the edge-on \hst-dark ALMA galaxies AID3 and AID5, right where the dust emission originates.
Thus, inclination likely does play a role in creating heavy dust obscuration in some of the \hst-dark galaxies \citep{nelson2023}.
Interestingly, we also find that the dust emission is always located at the center of the NIRCam counterpart, regardless of their edge-on or face-on morphologies. 
This implies that such significantly dust-obscured star formation plays a crucial role in the formation of bulges and/or the evolution to compact quiescent galaxies in the high-redshift universe \citep[e.g.,][]{lilly1999, genzel2003, tacconi2008, hickox2012, toft2014, chen2015, ikarashi2015, simpson2015a, barro2016, hodge2016, fujimoto2017, fujimoto2018, gullberg2019}.

\subsubsection{\jwst-dark galaxies}
\label{sec:jwst-dark}

The presence of \hst-dark galaxies at $z\sim3-5$ underlines the potential challenges associated with identifying similarly dust-reddened objects at $z\sim6$ even with \jwst/NIRCam \citep{kokorev2023}.
Indeed, several reports exist of IRAC-dark galaxies, invisible even in \textit{Spitzer}/IRAC maps, out to $z\sim$6--7 \citep[e.g.,][]{fudamoto2021, fujimoto2022, fujimoto2023b}.
Even with less dust obscuration, similar challenges would naturally arise for low-mass, moderately dust-obscured galaxies at very high redshifts ($z\gtrsim9$).
Therefore, we also explore ``\jwst-dark'' ALMA galaxies by defining the ALMA sources with the NIRCam counterpart whose intrinsic magnitude (i.e., after lens correction) in F444W is fainter than 30.0~mag. 
Among our 69 ALMA sources, we identify two such ALMA sources, AID60 and AID66, satisfying this criterion.

\begin{figure}
\includegraphics[trim=0cm 0cm 0cm 0cm, clip, angle=0,width=0.48\textwidth]{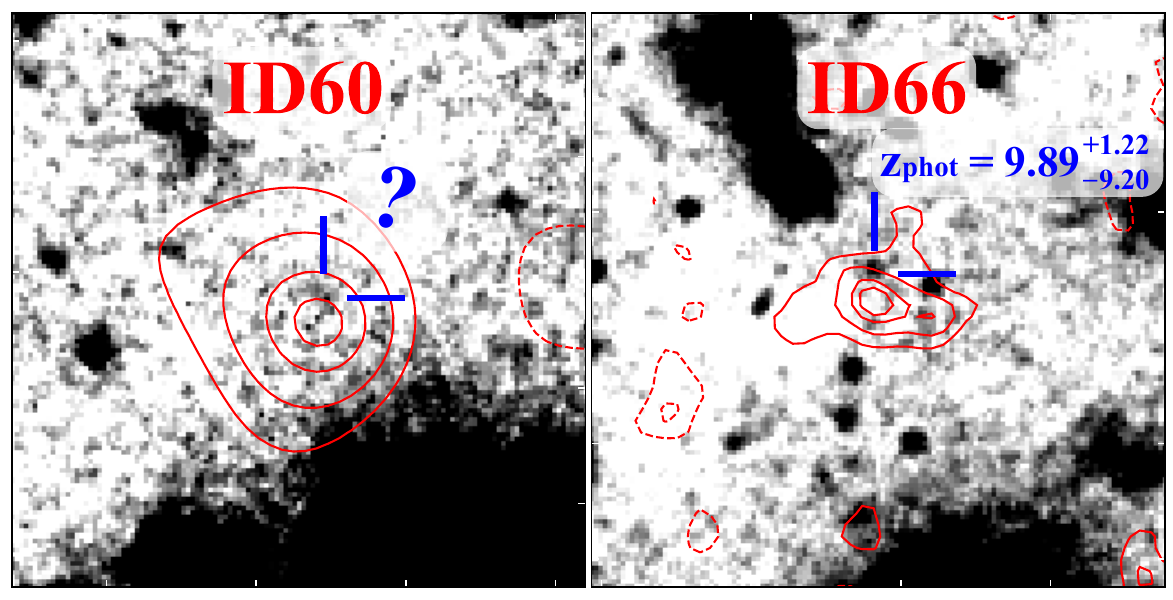}
\vspace{-0.4cm}
 \caption{
NIRCam F277W+F356W+F444W detection image cutout ($5''\times5''$) around the \jwst-dark (F444W $>$ 30~mag) ALMA galaxy candidates.
The red solid (dashed) contours represent the $3.0\sigma$, $3.5\sigma$, $4.0\sigma$, and $4.5\sigma$ ($-3.0\sigma$ and $-3.5\sigma$) significance levels of the ALMA 1.2-mm continuum in the \wide-natural and deep-tapered maps for AID60 and AID66, respectively, where the map showing the higher SNR is adopted.
The blue cross denotes the potential NIRCam counterpart, with that of AID66 estimated to have $z_{\rm phot}\sim9$ by our SED fits \citep{bwang2024}, but that of AID60 is not included in the NIRCam catalog \citep{weaver2023}, placing a 2$\sigma$ upper limit of 29.4mag in F444W. 
While the ALMA SNR close to the detection limit indicates these could be just spurious, the probability of chance projection of $z>9$ NIRCam candidates is estimated to be $\sim0.04$\% (see text).  
\label{fig:jwst-dark}}
\end{figure}

Figure~\ref{fig:jwst-dark} shows the ALMA contours overlaid on the combined F277W+F356W+F444W image. Continuous positive pixels are found near both AID60 and AID66 within a radius of $0\farcs25$. The positional uncertainty for these two ALMA sources is estimated to be $\sim0\farcs2-0\farcs3$ considering their beam sizes and SNRs\footnote{\url{https://help.almascience.org/kb/articles/what-is-the-absolute-astrometric-accuracy-of-alma}}, suggesting these continuous pixels may represent the NIRCam counterparts for these ALMA sources. The potential counterpart of AID66 is included in the catalog of \cite{weaver2023}, while that of AID60 is not. This places a 2$\sigma$ upper limit with a $0\farcs32$-aperture of 29.4~mag in F444W (30.1~mag after the lens correction, assuming $z=9$).
For AID66, the SED fitting result suggests $z_{\rm phot}=9.89^{+1.22}_{-9.20}$, 
with a consistent $z_{\rm phot}$ estimate also obtained via \eazy\ (see Table~\ref{tab:catalog}).
The lower-$z$ solution in the $z_{\rm phot}$ estimate is a natural consequence of the lack of secure upper limits bluer than the Lyman-$\alpha$ break, even with deep NIRCam maps, for faint sources of this level.
We confirm that similar NIRCam (F444W $>30$~mag) and ALMA ($S_{\rm 1.2mm}\sim0.2$~mJy) SED properties are reproduced by shifting the best-fit IRAC-dark galaxy template of REBELS-29-2 \citep{fudamoto2021} to $z\geq9$, indicating that these \jwst-dark ALMA galaxy candidates could be higher-redshift versions of the \hst-dark and IRAC-dark galaxies.
Such low-mass, dusty galaxies at $z\geq9$ may represent the dust-obscured phase of the progenitors of the remarkably UV-bright galaxies recently discovered at $z>9$ \citep[e.g.,][]{carniani2024}, during which most of their stars are formed \citep{ferrara2024}.

\begin{figure*}[t!]
\includegraphics[trim=0cm 0cm 0cm 0cm, clip, angle=0,width=1.\textwidth]{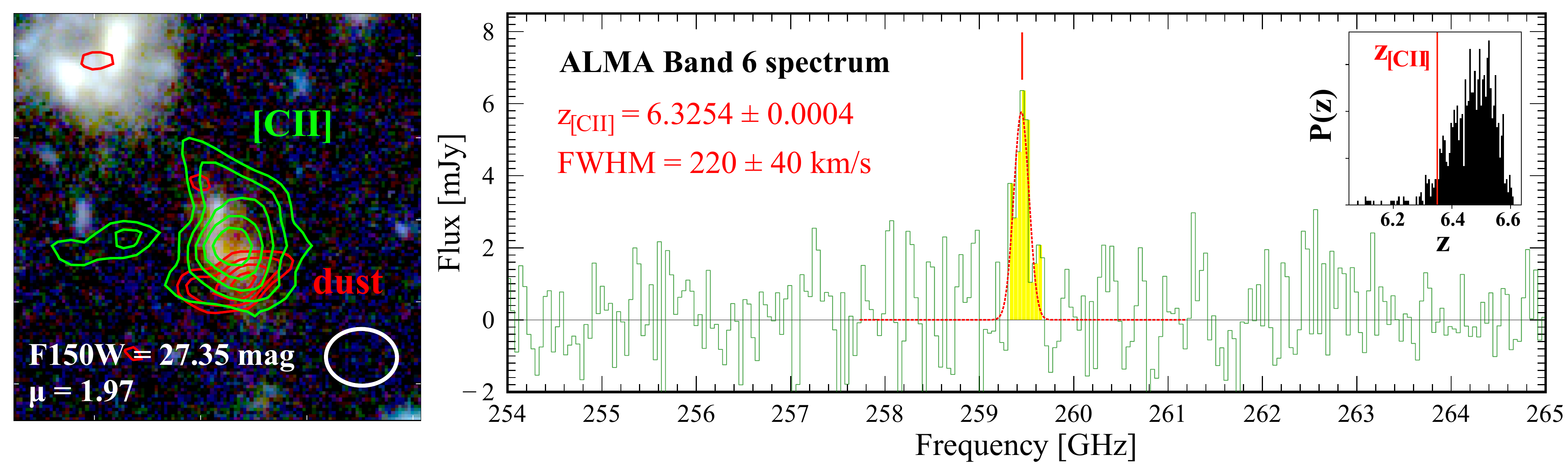}
\vspace{-0.4cm}
 \caption{
\cii\ line emitter at $z=6.33$ serendipitously detected (SNR $=7.0$ with an aperture) in the ALMA data cube of \survey.
\textit{\textbf{Left:}}
RGB NIRCam color cutout ($5''\times5''$) centered on the \cii\ emitter.
The green (red) contours indicate the $2\sigma$, $3\sigma$, $4\sigma$, $5\sigma$, and $6\sigma$ ($2\sigma$, $2.5\sigma$, $3\sigma$, and $3.5\sigma$) significance levels of the velocity-integrated \cii\ (dust continuum) intensity.
The white ellipse denotes the ALMA synthesized beam.
In the NIRCam map, three components are detected within a $0\farcs5$ radius, with $z_{\rm phot}$ values for all three components within $z\sim5.9$--6.6.
The inset labels show the properties of the nearest NIRCam counterpart.
\textit{\textbf{Right:}}
ALMA Band6 spectrum for the \cii\ emitter with the \deep-natural cube.
The red dashed curve presents the best-fit Gaussian for the line emission, and the inset label denotes the best-fit line properties. 
The inset panel shows the comparison between $z_{\rm [CII]}$ and $P(z)$ obtained from the \prospector\ fit for the nearest NIRCam counterpart. 
\label{fig:cii_emitter}}
\end{figure*}

\begin{figure*}[t]
\includegraphics[trim=0cm 0cm 0cm 0cm, clip, angle=0,width=1.\textwidth]{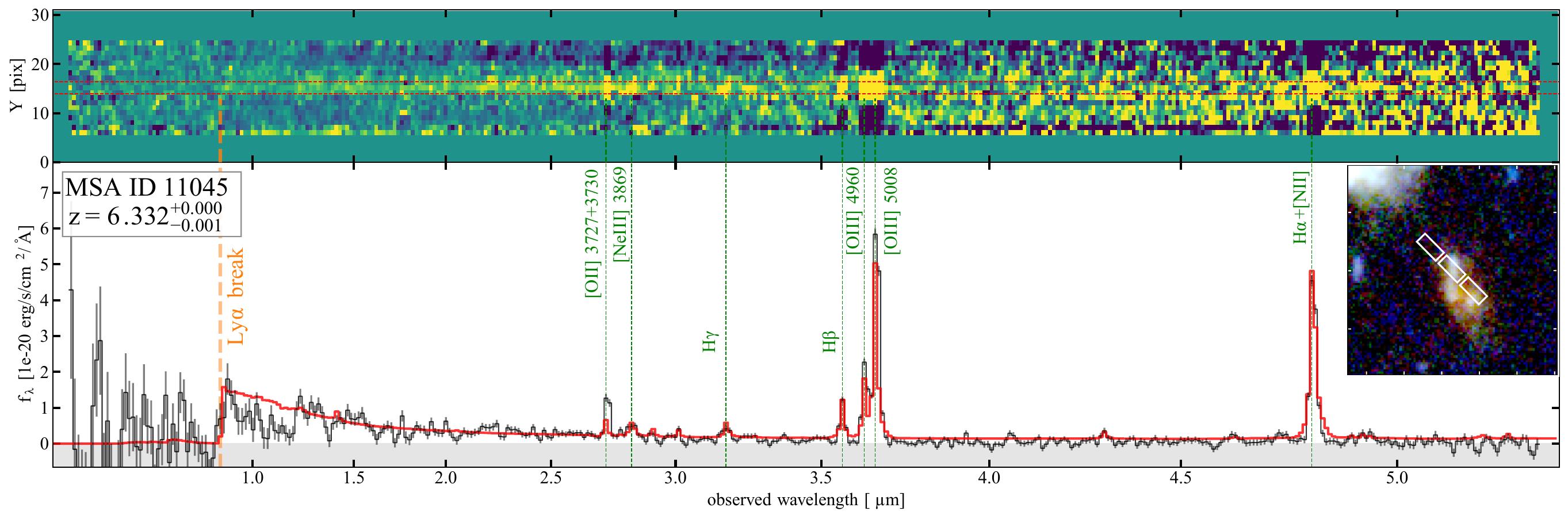}
\vspace{-0.4cm}
 \caption{
2D (\textit{top}) and 1D (\textit{bottom}) \jwst/NIRSpec prism spectra taken for the \cii\ line emitter at $z=6.33$. 
The inset panel shows the NIRCam RBG $3.6''\times3.6''$ cutout, where the three shutter positions are overlaid. 
The inset label shows its MSA ID and our best-fit redshift estimate from the prism spectrum. 
The red horizontal dashed lines in the top panel indicate the extraction aperture for the 1D spectrum.  
The black and red curves in the bottom panel denote the extracted 1D spectra and the best-fit {\sc eazy} template (see text). 
The green vertical lines represent wavelengths of the emission lines detected at SNR $\geq$ 4.0.
\label{fig:nirspec_z6}}
\end{figure*}

We note that AID60 and AID66 are detected in ALMA with SNR $=4.63$ and $4.93$, respectively, in the \wide-tapered and \deep-natural maps, whose selection thresholds are SNR $=4.4$ and 4.8 (Section \ref{sec:source_ext}). Given that one negative peak is detected close to the SNR threshold in each map (see Figure~\ref{fig:pn_hist}), their SNRs close to the selection thresholds suggest the possibility that both ALMA sources could be spurious.
Nonetheless, identifying a very faint $z_{\rm phot}\sim9$ NIRCam source close to a spurious source position is a remarkable coincidence. From the SED catalog of \cite{bwang2024}, we obtain the number density of $1.7\times10^{-3}$~arcsec$^{-2}$ for sources with F150W magnitudes comparable to the NIRCam counterpart of AID66 and $z_{\rm phot}$ estimates by our SED fit exceeding 8.5. This suggests that the probability of chance projection \citep{downes1986} with a $0\farcs25$ offset is approximately 0.03\%. 
If we examine the surface density of all NIRCam sources with the comparable F150W magnitude ($>$29.5~mag), regardless of the redshift, this probability increases to $\sim6$\%. 
We also note that AID60 meets our \jwst-dark criterion of F444W$>30$~mag after applying a lensing correction of $\mu=2.2$ assuming $z=9$. We confirm that the magnification remains $\mu=2.0$ when assuming $z=2$, indicating that AID60 is classified as a \jwst-dark source regardless of the redshift solution.

In summary, while we cannot rule out the possibility that these ALMA sources are spurious, these results suggest that AID60 and AID66 may be plausible \jwst-dark ALMA galaxy candidates, representing low-mass, moderately dusty galaxies at $z\sim9$ which may represent the dusty obscured-phase of the progenitors of the remarkably UV-bright galaxies recently discovered at $z>9$ \citep{ferrara2024}.

\subsection{\cii\ line emitter at $z=6.33$}
\label{sec:cii_emitter}

The homogeneous ALMA Band~6 mapping with the 30-GHz-wide ($\sim$244--274~GHz) frequency setup offers us a unique opportunity to search for line emitters in a blind manner. In particular, the frequency coverage corresponds to the \cii\ 158$\mu$m line redshift from 5.94 to 6.79. In this Section, we search for the \cii\ line emitters screened by NIRCam priors. 

We produce ALMA Band~6 spectra with the beam-size aperture at source positions of 150 NIRCam galaxies that are bright (F150W $<$ 27.5~mag) with $z_{\rm phot}$ estimates at $z_{\rm phot}=$6--7 in both \prospector\ and \eazy. 
Figure~\ref{fig:cii_emitter} shows a plausible \cii\ line emitter found in our search among the ALMA Band~6 spectra. The line feature is detected at $259.445\pm 0.015$~GHz with the line width of $220\pm40$~km~s$^{-1}$, resulting in the \cii-based redshift estimate of $z=6.3254\pm0.0004$.   
The NIRCam segmentation map suggests three components around the \cii\ source positions within $0\farcs5$, indicative of a merging system. 
The $z_{\rm phot}$ estimates of all three components are consistent within $z\sim5.9$--6.6, which is precisely consistent with the \cii-based redshift (see $P(z)$ in Figure~\ref{fig:cii_emitter}). 
The velocity-integrated map shows the \cii\ SNR $=6.3$ at the peak pixel with an elongated shape that matches the morphology observed in the NIRCam maps. With aperture-based photometry optimized to the \cii\ line structure, the \cii\ SNR increases to 7.0, suggesting that this is a comparably secure identification of the \cii\ line emitters presented in previous ALMA studies in a blind manner \citep[e.g.,][]{fujimoto2021}. The magnification factor of the NIRCam counterpart is estimated to be $\mu=1.97$. We infer the \cii\ luminosity to be $\log(L_{\rm [CII]}/L_{\odot})=9.1$ in the observed frame (i.e., without lens correction). 
The dust continuum is also detected at SNR $\sim4$\footnote{
While our continuum map was generated without masking the \cii\ line channels, the contribution of the \cii\ line flux within the 30-GHz bandwidth is estimated to be $\sim$30~$\mu$Jy, which is comparable to the 1$\sigma$ noise level. Also given the offset between \cii\ and dust, the \cii\ line has a negligible effect on the dust continuum detection.
}

, while this is not included in the 69 ALMA continuum sources due to the SNR thresholds adopted in the initial blind search  (Section~\ref{sec:source_ext}). This continuum detection suggests the infrared luminosity of $\log(L_{\rm IR}/L_{\odot})=11.6$ in the observed frame, assuming a modified blackbody with a dust temperature of 45~K and dust emissivity index of 1.8 \citep[e.g.,][]{liang2019, sommovigo2022}. 

\begin{table}
\setlength{\tabcolsep}{22pt}
\begin{center}
\caption{Observed physical properties of the \cii\ line emitter at $z=6.33$}
\vspace{-0.4cm}
\label{tab:cii_prop}
\begin{tabular}{lc}
\hline 
\hline
RA       [deg]    & 3.6123908    \\ 
Decl.    [deg]    & $-$30.4056401  \\ 
$z_{\rm [CII]}$   &  $6.3254\pm0.0004$   \\
$z_{\rm prism}$ &  $6.332\pm0.001^{\dagger}$   \\
$m_{\rm F150W}$ [mag]  & $27.35 \pm 0.09$ \\ 
$m_{\rm F444W}$ [mag]  & $26.35 \pm 0.04$  \\
$\mu$             &  1.97   \\
FWHM (\cii) [km~s$^{-1}$] & $220 \pm 40$ \\ 
$L_{\rm [CII]}\times \mu$ [$10^{9}\, L_{\odot}$]  &     $1.37\pm0.22$   \\
$f_{\rm 1.2mm}\times \mu$ [$\mu$Jy]  &     $130 \pm 32$   \\
$L_{\rm IR}\times \mu$ [$10^{11}\, L_{\odot}$]  &     $4.02\pm0.99^{\sharp}$   \\
$L_{\rm H\alpha+[NII]}\times \mu$  [$10^{8}\, L_{\odot}$]  & $10.8\pm 0.3$ \\
$L_{\rm H\beta}\times \mu$  [$10^{8}\, L_{\odot}$]  & $2.74 \pm 0.20$ \\
$L_{\rm H\gamma}\times \mu$  [$10^{8}\, L_{\odot}$]  & $1.10 \pm 0.22$ \\
$L_{\rm [OIII]5008}\times \mu$  [$10^{8}\, L_{\odot}$]  & $16.3\pm0.3$ \\
$L_{\rm [OII]3727+3730}\times \mu$  [$10^{8}\, L_{\odot}$]  & $3.74 \pm 0.32$ \\
$L_{\rm [NeIII]3869}\times \mu$  [$10^{8}\, L_{\odot}$]  & $1.23\pm 0.30$ \\
R23 & $9.28\pm1.06$ \\
O32 & $4.35\pm0.38$ \\
Ne3O2 &  $0.33\pm0.09$ \\
12+log(O/H) & $7.84_{-0.16}^{+0.25}$ \\
$A_{\rm v}$ [mag] & $1.40 \pm 1.40^{\natural}$                    \\ 
SFR$_{\rm H\beta}\times \mu$ [$M_{\odot}$ yr$^{-1}$] &  67$^{+223\flat}_{-52}$   \\
$M_{\rm star}\times \mu$  [$10^{8}\,M_{\odot}$] & $3.4^{+1.7}_{-1.3}$  \\ \hline
MSA ID     &     11045        \\ 
NIRCam ID$^{\partial}$  &     12053     \\ \hline\hline
\end{tabular}
\end{center}
\tablecomments{
The coordinate denotes the \cii\ line peak position in the velocity-integrated map. 
The physical properties based on the NIRCam and NIRSpec data represent one of the nearest NIRCam counterparts, which is observed with the NIRSpec prism (Figure~\ref{fig:nirspec_z6}). \\
$\dagger$ The slight redshift difference may indicate the velocity difference between the \cii-emitting and the NIRSpec-observed regions. \\
$\sharp$ Assuming a modified black body with a dust temperature of 45~K and dust emissivity index of 1.8. \\
$\natural$ Based on the Balmer decrement via H$\gamma$/H$\beta$. \\
$\flat$ Using the calibration of \cite{murphy2011}, after dust correction. 
The error bar is based on the uncertainty of $A_{\rm v}$. \\
$^{\partial}$ Source ID in the DR2 UNCOVER NIRCam Source catalog of \cite{weaver2023}. 
}
\end{table}

One of the three NIRCam components was assigned in an MSA mask of NIRSpec and received 2.7 hours of exposure with prism. 
In Figure~\ref{fig:nirspec_z6}, we show the 2D and 1D spectra taken for the NIRCam counterpart of the \cii\ line emitter. 
The inset panel shows the three shutter positions overlaid on the NIRcam RGB image. 
In the same fitting procedures as AID3 described in Section~\ref{sec:nirspec}, the source redshift is securely confirmed at $z=6.332$ with successful multiple line detection of \oiii5008, 4960, \oii3727+3730, H$\alpha$+\nii, H$\beta$, and H$\gamma$ at SNR $\geq4$. 
The dust attenuation is estimated via the Balmer decrement of H$\gamma$/H$\beta$ with the \cite{calzetti2000} law, resulting in $A_{\rm V}=1.4$, which infers the H$\beta$-based SFR after dust and lens correction of 34~$M_{\odot}$~yr$^{-1}$, using the calibration of \cite{murphy2011}. 
From the optical line ratios of (\oiii+\oii)/H$\beta$ ($\equiv$R23) and \oiii/\oii ($\equiv$O32), the gas-phase metallicity is also estimated to be 12+$\log$(O/H) = 7.84, with the calibration of \cite{nakajima2022}\footnote{To cover the parameter space of the relatively high R23 = 9.28 observed in the \cii\ line emitter, we use the R23-metallicity relation calibrated for large H$\beta$ equivalent width sources in \cite{nakajima2022}. In this conversion, we assume 12+$\log$(O/H)$\gtrsim8.1$, which is implied from the high O32 (=4.35) and the O32-metallicity calibration \citep[e.g.,][]{maiolino2008, curti2017, bian2018, nakajima2022}}. 
Based on these SFR and metallicity measurements, the \cite{vallini2015} model predicts the \cii\ line luminosity from this source of $\log(L_{\rm [CII]}/L_{\odot})=8.6$. 
Although the \cii\ luminosity obtained from the observation ($\log(L_{\rm [CII]}/L_{\odot})=9.1$) is higher than this model prediction, we confirm that this difference is still consistent within the 1$\sigma$ error, after taking the uncertainties in the metallicity estimate and the dust correction for the SFR estimate into account. 
We also confirm that the SFR-\cii\ relation calibrated with local metal-poor dwarf galaxies in \cite{delooze2014} suggests $\log(L_{\rm [CII]}/L_{\odot})=9.3$ with our SFR estimate. 
We thus conclude that the \cii\ line emitter at $z=6.33$ found in \survey\ falls in the typical SFR--$L_{\rm [CII]}$ relation within the uncertainties of both measurement and relation, instead of being an exceptionally \cii\ bright source. 
The ratio between \cii\ and IR luminosity, $\log(L_{\rm [CII]}/L_{\rm IR})=-2.5$,  is comparable to those found in star-forming galaxies of similar $L_{\rm IF}$ at low redshift. It exceeds the typical luminosity ratio seen in observations at $z>5$ by a factor of a few (see e.g., \citealt{liang2023}). However, it is still consistent with previous results given the large observed scatter of the line ratio in e.g., REBELS \citep{bouwens2022aa}. 
In Table~\ref{tab:cii_prop}, we summarize our measurements and uncertainties for the physical properties of the \cii\ line emitter at $z=6.33$.

We note a slight redshift difference ($\approx300$~km~s$^{-1}$) is observed between the ALMA- and NIRSpec-based redshift measurements. 
This offset is consistent with the mean offset value recently reported in the NIRSpec prism spectra due to the intrashutter source positioning effect \citep{deugenio2025}, and thus it may be attributed by the uncertainty of the wavelength calibration.
Another possibility is the slight positional difference in the \cii-emitting and NIRSpec-observed regions (see Figure~\ref{fig:cii_emitter} and \ref{fig:nirspec_z6}), where a complex gas kinematic may take place due to the merging process. 
Therefore, we remark on this potential spatial offset effect for readers to also consider when comparing the physical properties constrained with ALMA and NIRSpec for the \cii\ line emitter.

Owing to the blind aspect of the \survey\ survey, our successful identification of the \cii\ line emitter enables us to provide a lower limit on the \cii\ luminosity function (LF) at $z=6-7$. Assuming a typical line width of FWHM = 200~km~s$^{-1}$ and a 5$\sigma$ detection limit, we obtain an effective survey area after the lens correction of $\sim2.3$~arcmin$^{2}$ at $\log(L_{\rm [CII]}/L_{\odot})=8.8$. Based on the redshift range of $z=$5.94--6.79 and the lower limit estimate from the Poisson uncertainty at the single-sided confidence level of 84.13\% presented in \cite{gehrels1986}, we infer the lower limit of $3.2\times10^{-5}$~Mpc$^{-3}$.

\begin{figure}
\includegraphics[trim=0cm 0cm 0cm 0cm, clip, angle=0,width=0.48\textwidth]{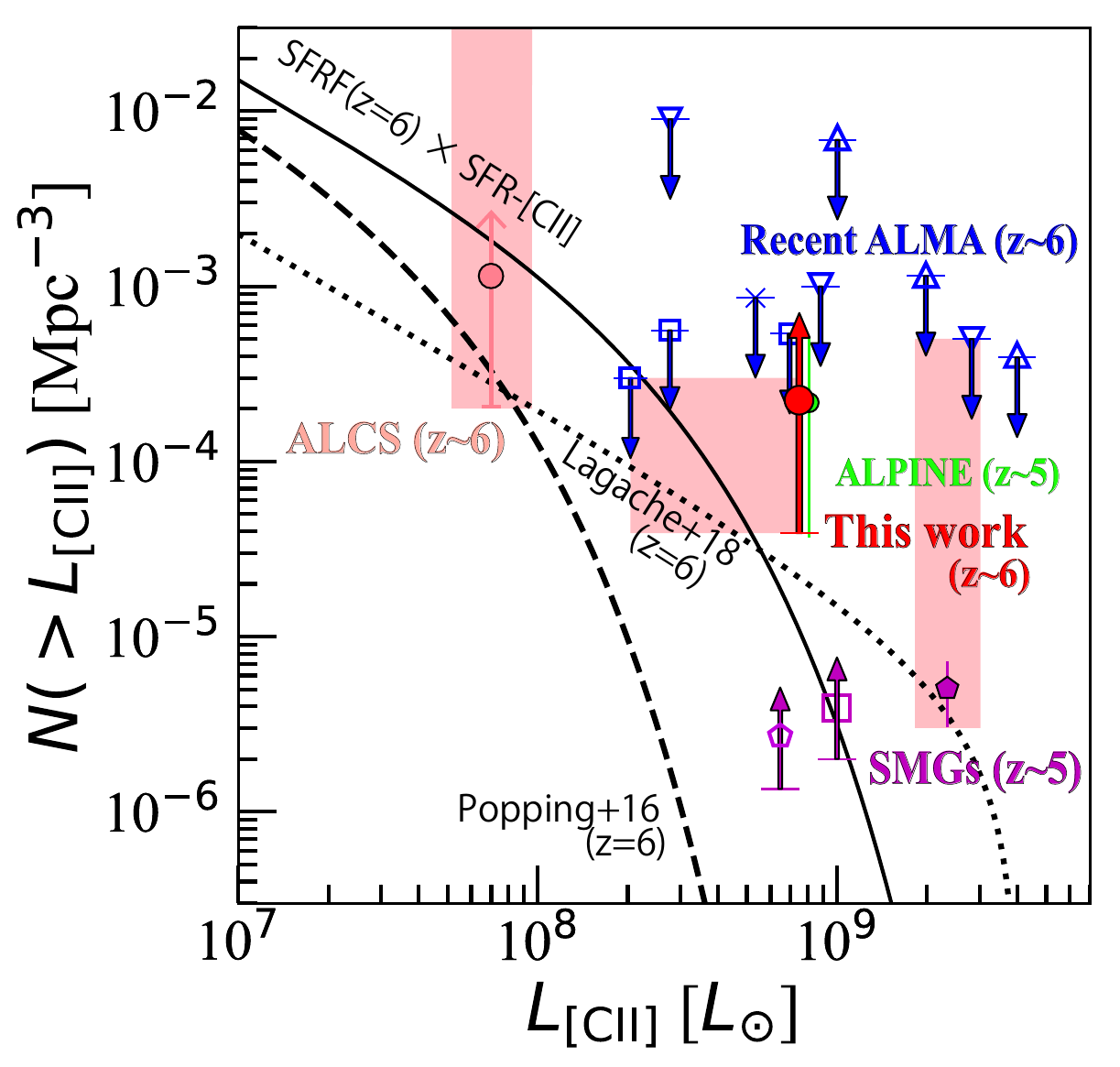}
\vspace{-0.4cm}
 \caption{
Cumulative \cii\ luminosity function from recent \cii\ line studies at $z\sim5$--6. The red circle shows the volume density of the \cii\ line emitter, based on our single successful detection in \survey\ (Section~\ref{sec:cii_emitter}).
The lower limit is estimated from the Poisson uncertainty at the single-sided confidence level of 84.13\%, as presented in \cite{gehrels1986}.
Recent ALMA blind line survey results are indicated by the blue triangle (243 archival
data cubes; Matsuda et al. 2015), blue inverse triangle (four lensing clusters; \citealt{yamaguchi2017}), blue square (ASPECS; \citealt{decarli2020, uzgil2021}), 
and blue cross (SSA22; \citealt{hayatsu2017, hayatsu2019}), respectively. 
The green circle denotes the ALPINE results \citep{loiacono2020}. 
The magenta square and pentagon show the serendipitous \cii\ line detection from bright SMGs at $z\sim5$ reported in \cite{swinbank2012} and \cite{cooke2018}, respectively. 
The red-shaded regions indicate the current constraints based on both our results and those previously obtained. For comparison, we also present the semi-analytical model results \citep{popping2016, lagache2018} and the SFR function (SFRF; \citealt{smit2018}), including the dust correction, whose SFR value is converted into $L_{\rm [CII]}$ using the local empirical relation \citep{delooze2014}.
\label{fig:cii_lf}}
\end{figure}

\setlength{\tabcolsep}{10pt}
\begin{deluxetable}{ccc}
\tablecaption{Constraints on \cii\ Luminosity Function at $z\sim6$}
\tablehead{\colhead{$\log(L_{\rm [CII]}/L_{\odot})$}  & \colhead{$\log(\Phi)$ (lower)}  & \colhead{$\log(\Phi)$ (upper)} \\
    &       [Mpc$^{-3}$~dex$^{-1}$]      &   [Mpc$^{-3}$~dex$^{-1}$]}
\startdata
7.85 & $-3.71$ & \nodata \\ 
8.28--8.84 & $-4.49$ & $-3.47$ \\
9.37--9.45 & $-5.52$ & $-3.30$ \\
\enddata
\tablecomments{
Lower and upper boundaries drawn as the red shaded regions in Figure~\ref{fig:cii_lf} constrained from \survey\ and previous ALMA studies. 
}
\label{tab:cii_lf}
\end{deluxetable}

In Figure~\ref{fig:cii_lf}, we present the cumulative volume density of the \cii\ line emitters, including our lower limit estimate and recent ALMA measurements at $z\sim5$--6 \citep{swinbank2012, matsuda2015, yamaguchi2017, cooke2018, hayatsu2019, decarli2020, yan2020, fujimoto2021, uzgil2021}. The red shaded regions indicate the possible parameter space from our and recent ALMA measurements. For comparison, we also show predictions from the semi-analytical models \citep{popping2016, lagache2018} and from the empirical relations by combining the observed SFR function (SFRF; \citealt{smit2016}) for optically-selected galaxies and the SFR--$L_{\rm [CII]}$ relation calibrated among local star-forming galaxies \citep{delooze2014}. 
We find the predictions from both semi-analytical models and the SFRF falling below our lower limit estimate, indicating that the \cii\ emitters could be more abundant than the predictions from the recent galaxy models. The prediction from the SFRF also falls below the measurement from SMGs at the bright end. This is likely explained by the lack of the heavily dust-obscured galaxies in the optically-selected galaxies that are used for the SFRF measurement \citep{smit2016}.

\subsection{ALMA views of $z\gtrsim9$ galaxies}
\label{sec:z9}

\begin{figure*}[t!]
\includegraphics[trim=0cm 0cm 0cm 0cm, clip, angle=0,width=1.\textwidth]{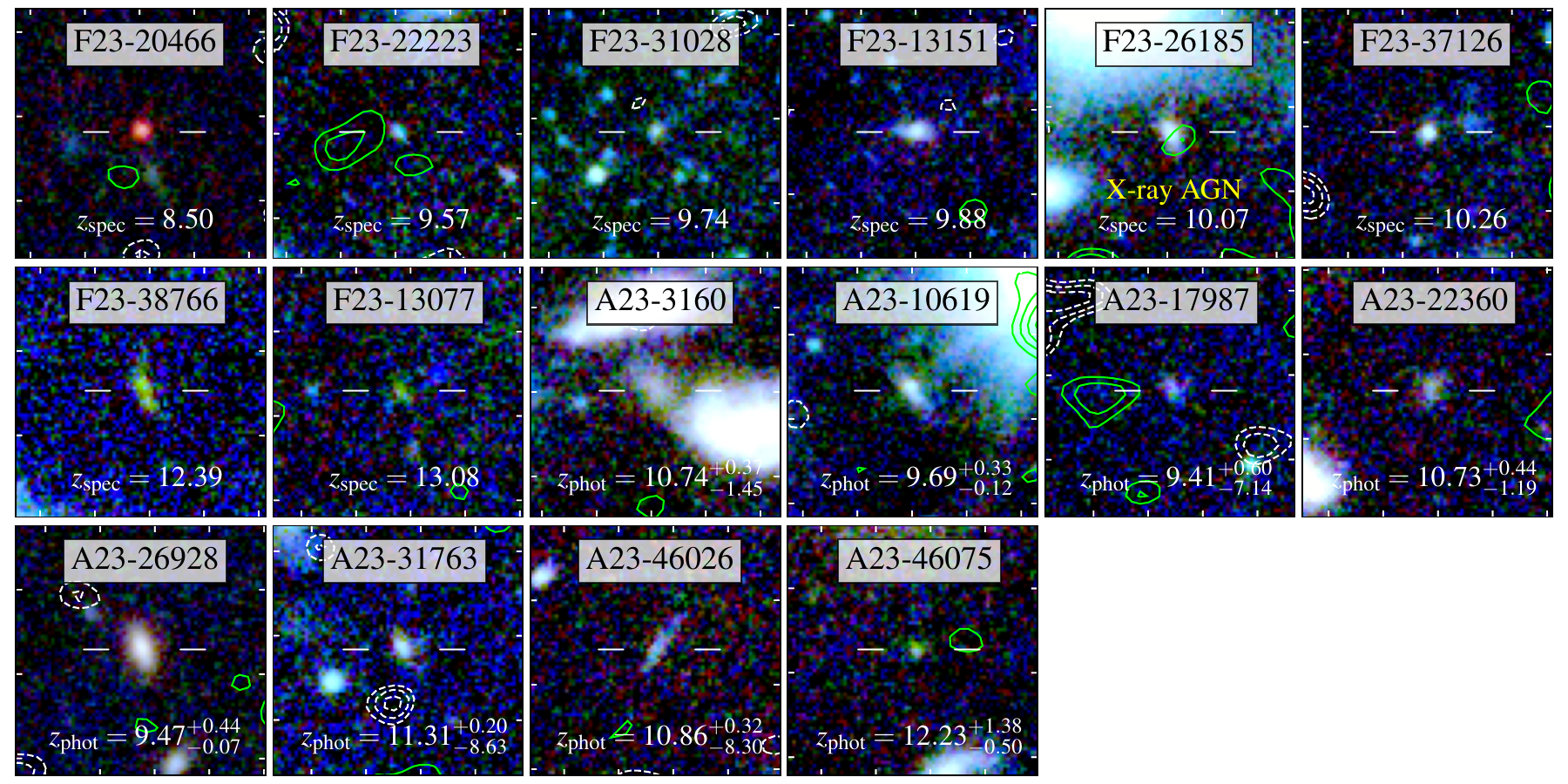}
\vspace{-0.2cm}
 \caption{
ALMA 1.2-mm views of $z\gtrsim9$ galaxies in UNCOVER.  
RGB NIRCam color cutouts ($3''\times3''$) for spec-$z$ confirmed galaxies presented in \cite{fujimoto2023c} (hereafter F23) and photometric galaxies presented in \cite{atek2023} (hereafter A23) at $z\gtrsim9$. 
The white bars remark the $z\gtrsim9$ galaxies, 
and the inset labels present their IDs and their $z_{\rm spec}$ or $z_{\rm phot}$ estimates in F23 or A23. 
The green (white) contours denote $2.0\sigma$, $2.5\sigma$, and $3.0\sigma$ ($-2.0\sigma$, $-2.5\sigma$, and $-3.0\sigma$) significance levels of the ALMA 1.2-mm continuum in the natural map. 
F23-26185 has been reported to be an X-ray AGN at $z\sim10$ \citep{bogdan2023} and confirmed at $z=10.07$ (see also \citealt{goulding2023}). 
Interestingly, the marginal ALMA detection ($\sim2.6\sigma$) coinciding with the source position is identified only in F23-26185, implying an early active co-evolution of the central black hole and its host at $z>10$.  
\label{fig:z9}}
\end{figure*}

\begin{figure}
\includegraphics[trim=0cm 0cm 0cm 0cm, clip, angle=0,width=0.5\textwidth]{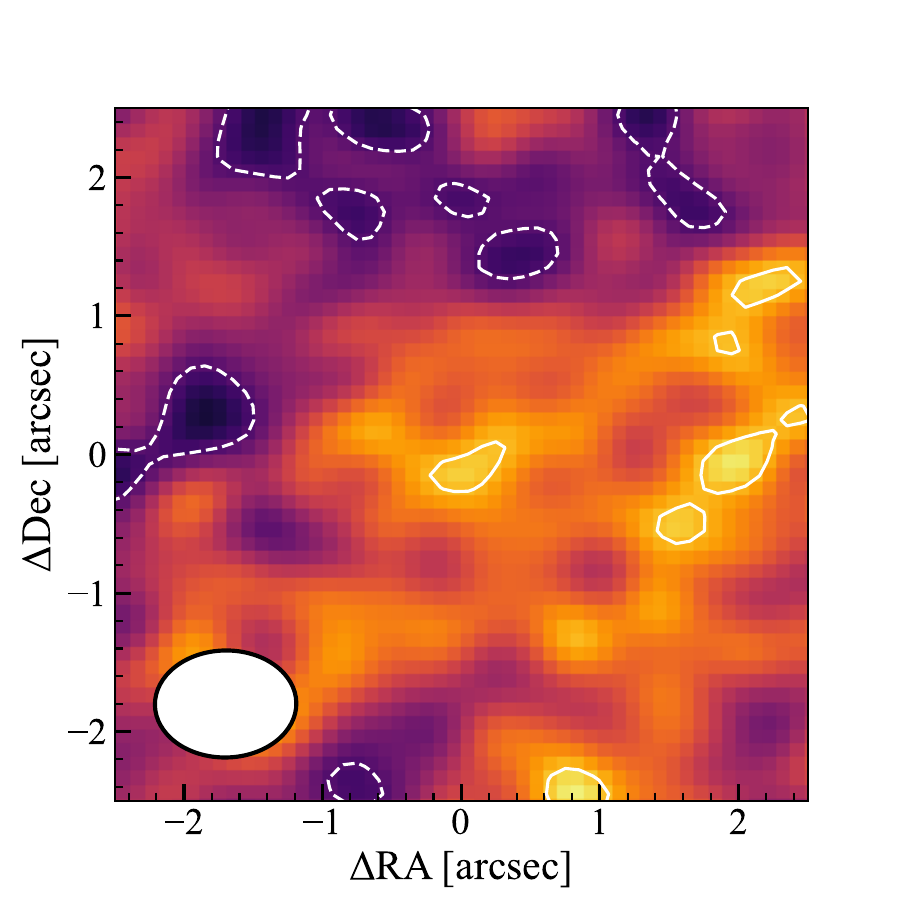}
\vspace{-0.2cm}
 \caption{
Stacked ALMA 1.2-mm continuum map constructed using the \wide-natural images of 16 galaxies with $z_{\rm spec}$ or $z_{\rm phot}\gtrsim9$ presented in Figure~\ref{fig:z9}. Solid (positive) and dashed (negative) contours start at $\pm2\sigma$ and increase in steps of $1\sigma$, where $1\sigma=11.8\,\mu{\rm Jy\,beam}^{-1}$ after primary beam correction. The synthesized beam is shown as the white ellipse at the bottom left. A marginal detection ($\sim2.3\sigma$) is visible at the central stacked position, implying an intrinsic (lens-corrected) flux density of $14.0\pm6.2\,\mu{\rm Jy}$ when adopting a median magnification factor of 1.9 for the stacked sources.
\label{fig:z9_stack}}
\end{figure}

Recent ALMA observations report a sizeable fraction of bright optically-selected galaxies at $z\sim4$--8 have dust emission similar to those of low-$z$ starbursts \citep[e.g.,][]{watson2015, bowler2018, tamura2019, bethermin2020, inami2022, witstok2022}, indicating that the dust-obscured star formation may still contribute to a notable portion of the total SFR density out to $z\sim7$ \citep{fujimoto2023b, algera2023, barrufet2023b}. 
Following the successful identifications of dozens of bright galaxy candidates at $z\gtrsim9$ with \jwst\ \citep[e.g.,][]{castellano2022, naidu2022c, donnan2023, harikane2023, finkelstein2023}, deep ALMA observations have also been swiftly performed through the Director's Discretionary Time (DDT) for several of these bright \jwst\ galaxy candidates at $z\sim9$--17. In these ALMA DDT observations, no robust dust continuum detection has been made so far \citep{fujimoto2022b, fujimoto2022c, bakx2023, yoon2023}, while a marginal ($\sim3\sigma$) detection is reported from the remarkably luminous galaxy candidate at $z_{\rm phot}=10.5$ \citep{yoon2023}. 

To enrich our understanding of the ALMA views of high-redshift galaxies newly identified with \jwst\ with a larger sample, we also investigate whether $z\gtrsim9$ galaxy candidates identified in the \survey\ field show any marginal dust continuum detection or not. 
In Figure~\ref{fig:z9}, we show the ALMA Band~6 1.2-mm continuum contours obtained from the \wide-natural map and overlaid on the NIRCam color $3''\times3''$ cutout for 8 galaxies that are spectroscopically confirmed at $z\geq8.5$ in the NIRSpec MSA follow-up in A2744 \citep{roberts-borsani2023, boyett2023, goulding2023, bwang2023, kokorev2023b, fujimoto2023c} and fall in the ALMA footprint of \survey. 
Because \cite{fujimoto2023c} report the 100\% success ratio among the photometric candidates selected in \cite{atek2023}, we also present 8 galaxy candidates at $z\simeq9$--15 presented in \cite{atek2023} that are not included in the MSA design, but have the same robustness at $z\gtrsim9$.
We find that none of these \jwst\ high-$z$ galaxies show a dust continuum above the $3\sigma$ level, while F23-26185 shows a marginal ALMA detection ($\sim2.6\sigma$) coinciding with the \jwst\ source position with the spatial offset of $\sim0\farcs1$. 
Interestingly, F23-26185 has been reported to show an X-ray detection in the 1.25~Ms deep \textit{Chandra} data, suggesting the emergence of a massive black hole ($M_{\rm BH}\approx4\times10^{7}M_{\odot}$; \citealt{bogdan2023}) at $z_{\rm phot}=10.3$ (UHZ1; e.g., \citealt{castellano2023}), and the follow-up NIRSpec spectroscopy has successfully confirmed its redshift at $z=10.07$ (see also \citealt{goulding2023}). 
Although it is difficult to draw a definitive conclusion as to whether the marginal ALMA detection is really associated with F23-ID26185 or just spurious with this low significance of the ALMA emission, it is an interesting coincidence that such a marginal detection only happens in the $z=10.07$ X-ray AGN among the 16 \jwst\ sources at $z\simeq9$--15. 
If it is real, 
the inferred obscured SFR fraction from this potential dust continuum emission is $\approx50\%$, where comparably highly obscured $z\gtrsim7$ galaxies have been recently reported \citep[e.g.,][]{akins2022,algera2023}.
The marginal ALMA detection may indicate the dusty star-forming activity in the host galaxy, which would play an important role in the early co-evolution between the central super-massive black hole and its host \citep[e.g.,][]{wang2013, willott2015, venemans2018, izumi2019, pensabene2020, neeleman2021, fujimoto2022}.

In the \wide-natural map, we identify 3388 sources with SNR = 2.5--3.0, yielding the number density of such marginal signals to be $3.9\times10^{-2}$~arcsec$^{-2}$ in our ALMA map. This infers the probability of the chance projection \citep{downes1986} with the $0\farcs1$ offset to be $\sim$0.77\%. 
While these calculations indicate that the probability of the marginal ALMA detection being spurious still remains, the possibility decreases by multiplying with probabilities of facts that similar marginal detections are not identified in all other 15 galaxies at $z\gtrsim9$, but identified only in the $z=10.07$ X-ray AGN host galaxy.   

In short, although there remains a small likelihood ($<1\%$) of the signal being spurious, the marginal ALMA detection in F23-26185 (a.k.a UHZ1) is a plausible signature of the presence of the active co-evolution of the central black hole and its host at $z>10$. 

In~Figure~\ref{fig:z9_stack}, we present a stacked ALMA map generated from the \wide-natural images of 16 sources at $z_{\rm spec}$ or $z_{\rm phot}\gtrsim9$. The 1$\sigma$ noise level is reduced to 11.8~$\mu$Jy/beam after primary beam correction, corresponding to an effective noise of 6.2~$\mu$Jy/beam when accounting for lensing magnification (median magnification factor of $1.9$). We detect marginal ($\sim2.3\sigma$) emission at the stacked position, inferring an intrinsic (lens-corrected) 1.2-mm flux density of $14.0\pm6.2$~$\mu$Jy. As in the case of UHZ1, the current S/N remains insufficient for a definitive conclusion. Nevertheless, these results demonstrate the potential of deep ALMA observations with gravitational lensing to probe the faint dust continuum emission from galaxies at $z\gtrsim9$.

\subsection{First Look at IRLF out to $z\sim10$}
\label{sec:irlf}

\begin{figure*}[t!]
\includegraphics[trim=0cm 0cm 0cm 0cm, clip, angle=0,width=0.98\textwidth]{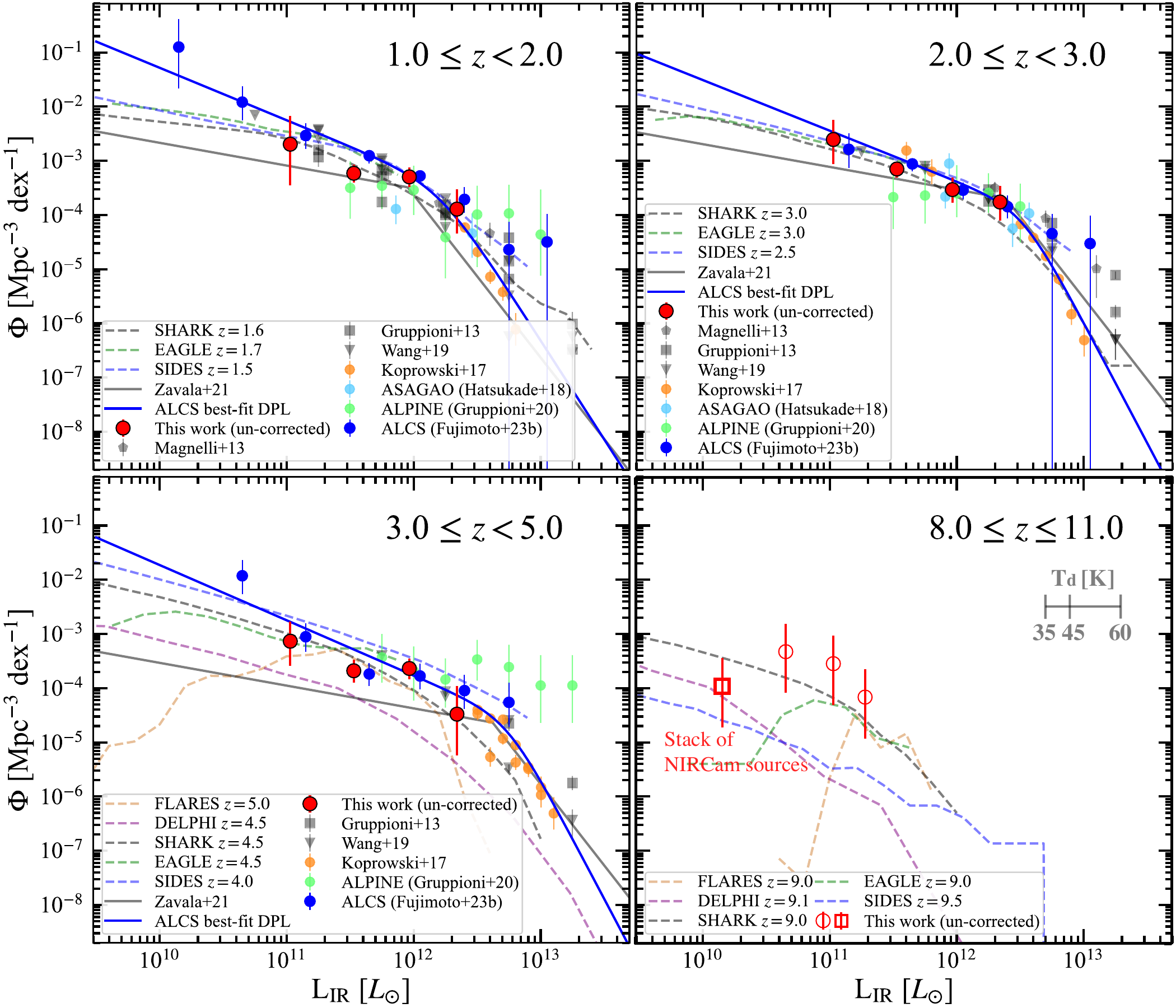}
\vspace{-0.2cm}
 \caption{
IR LF measurements out to $z\sim10$. 
The red-filled circles represent our \survey\ measurements.  
The red-open circles indicate the possible constraints from the two \jwst-dark ALMA galaxy candidates (Section~\ref{sec:jwst-dark}) and the marginal ($\sim2.6\sigma$) ALMA detection from the X-ray AGN at $z=10.07$ (Section~\ref{sec:z9}) if they are real. 
The red open square denotes the marginal ($\sim2.3\sigma$) ALMA detection in the stacked image of the 16 NIRCam sources at $z\gtrsim9$ presented in Figure~\ref{fig:z9}.
Our measurements do not include completeness correction.  
Previous IR LF measurements with \textit{Herschel} \citep{magnelli2013, gruppioni2013, wang2019} ALMA+SCUBA2 \citep{koprowski2017}, and ALMA \citep{hatsukade2018, dudzeviciute2020, gruppioni2020, fujimoto2023b} are shown in grey and other colored symbols, together with the best-fit DPL function estimated in \cite{zavala2021} and \cite{fujimoto2023b}. 
The other color lines show the predictions from simulations \citep{lagos2020, trayford2020, vijayan2022, bethermin2022, mauerhofer2023}, where the turnover in the faint IR luminosity range in some of these simulations is caused by the incompleteness due to the mass resolution limit in their calculations. 
The grey bar represents the increase of the $L_{\rm IR}$ at $z=9$ with increasing $T_{\rm d}$ assumption.
\label{fig:irlf}}
\end{figure*}

The first public homogeneous wide ALMA and \jwst\ blind maps of \survey\ provide us a new NIR-mm window to identify the high-redshift galaxy populations more comprehensively than ever before. 
The successful identification of the \hst-dark ALMA galaxies at $z\sim3-5$ (Section~\ref{sec:hst-dark}) indicates that \survey\ overcomes the incompleteness of the galaxy surveys at $<2$~$\mu$m wavelengths and directly measures their dust-obscured star-forming activities. 
The further potential identification of the \jwst-dark ALMA galaxies (Section~\ref{sec:jwst-dark}) and the marginal ALMA detection from the X-ray AGN at $z=10.07$ (Section~\ref{sec:z9}) also enable us to obtain first possible constraints on the dust-obscured activities at $z\gtrsim9$. In this Section, we evaluate the IRLFs at $z\sim1$--5 from our ALMA sources and at $z\gtrsim9$ from the two \jwst-dark ALMA galaxy candidates and the marginal ALMA detection in the X-ray AGN at $z=10.07$. 

Based on our $z_{\rm photo}$ estimates (Section~\ref{sec:phy_properties}), we derive the IRLFs in the same manner as \cite{fujimoto2023b}, 
although we note that we assume a modified black body for the $L_{\rm IR}$ estimate due to the lack of multiple FIR-band constraints, different from \cite{fujimoto2023b}. 
We assume the modified black body with the dust temperature of $T_{\rm d}=35$~K and the dust-emissivity index $\beta_{\rm d}=1.8$ \citep[e.g.,][]{mauerhofer2023}. 
We calculate the uncertainty from the Poisson error based on the values presented in \cite{gehrels1986} that are applicable in small-number statistics. Note that we do not apply the completeness correction to our IRLF measurements in this paper as initial results. 
A complete measurement, including the completeness correction based on the proper size measurements and uncertainties from the magnification, 1.2-mm flux, and $L_{\rm IR}$ measurements, will be presented in a separate paper.  

In Figure~\ref{fig:irlf}, we show our IRLF measurements at $1\leq z<2$, $2\leq z<3$, and $3\leq z<5$. For comparison, we also show recent IRLF measurements both from ALMA \citep{koprowski2017, hatsukade2018,dudzeviciute2020,gruppioni2020, fujimoto2023b} and single-dish observations \citep{magnelli2013, gruppioni2013, wang2019}, the best-fit double power-law (DPL) functions obtained from the ALCS survey \citep{fujimoto2023b} and a backward evolution model with 1.2-mm and 3-mm number count constraints available in the literature \citep{zavala2021}, and the theoretical model predictions \citep{lagos2020, trayford2020, bethermin2022, vijayan2022, mauerhofer2023}. 
Our measurements are consistent with the previous measurements and the model predictions within their scatters. Compared with the best-fit DPL function of ALCS, we find that the faintest data points in our measurements always fall below the ALCS DPL function. This is likely explained by the lack of completeness correction in our measurements, which can be significant, especially in the faint end. 
On the other hand, we find that most of our measurements, especially at faint regimes ($L_{\rm IR}\lesssim10^{11.5-12}$), fall above the best-fit DPL obtained in \cite{zavala2021}. This is likely because the 1.2-mm and 3-mm number count constraints available at that time and used for the fit in \cite{zavala2021} did not cover those faint regimes in the wide redshift range, which might lead to the faint-end slope being underestimated. 

Figure~\ref{fig:irlf} also presents the potential constraints on the IRLF at $z=8-11$. The three individual data points (red open circles) are obtained by assuming that the two \jwst-dark ALMA galaxy candidates and the marginal ALMA detection in the X-ray AGN at $z=10.07$ are all real. 
The additional data point (red open square) is derived from the marginal detection obtained by stacking the \wide-natural ALMA images of the 16 NIRCam galaxies at $z\gtrsim9$ (Section~\ref{sec:z9}).
Here we calculate the survey volume with the SNR threshold of 2.5 and the redshift range of $z=8$--11, where we assume the redshift of the \jwst-dark ALMA galaxy candidates at $z=9$. The $L_{\rm IR}$ values are estimated in the same manner as above. 
We find that these potential constraints are still consistent with the model predictions of FLARES, SHARK, and EAGLE within the errors. 
Although these potential constraints could be upper limits, given the remaining possibility that their ALMA detections are just spurious (Section~\ref{sec:jwst-dark} and Section~\ref{sec:z9}), this comparison result indicates that the identification of the \jwst-dark ALMA galaxies and/or the marginal ALMA detection from X-ray AGN at $z\gtrsim9$ within the survey volume of \survey\ are not in tension with predicted abundances. 
We note the completeness correction increases our volume density estimates, which may make the situation challenging. However, the spatial size of the galaxies at $z\gtrsim9$ is small (effective radius of $\lesssim0\farcs1$; \citealt{ono2022}) relative to the beam sizes in our ALMA maps ($\sim1''-2''$). The lensing magnifications are estimated to be moderate ($\mu\sim2-4$) among these three sources. Therefore, the impact of the completeness correction, including the lensing distortion, is likely modest among these three sources. 
We also note that significant uncertainties remain in the estimate of $L_{\rm IR}$. For instance, an increasing dust temperature ($T_{\rm d}$) trend toward higher redshift has been reported \citep[e.g.,][]{schreiber2018}. Extrapolating the best-fit redshift evolution model for $T_{\rm d}$ from \cite{sommovigo2022} suggests $T_{\rm d}$ of $\sim$60~K at $z\sim10$. Adopting $T_{\rm d}=60$K would increase the inferred $L_{\rm IR}$ values of all data points in the $z=8$--11 panel of Figure\ref{fig:irlf} by a factor of $\sim$4, which brings a tension between the individual data points and the current simulations. 
On the other hand, recent studies have also identified the presence of dusty star-forming galaxies with $T_{\rm d}\approx$30--35~K at $z\sim7$, consistent with the assumptions adopted in our analysis. 
Deeper follow-up observations in multiple ALMA Bands are essential to improve the accuracy of $L_{\rm IR}$ estimates and to robustly constrain the IRLFs out to $z\sim10$.

\section{Impact of ALMA $\times$ \jwst\ legacy A2744 field on other topics}
\label{sec:result}

In Section~\ref{sec:result}, we overview our initial results from the \survey\ survey. 
We anticipate that the legacy aspect of \survey\ will continue to generate more discoveries through wide use from the community, including the discovery we cannot fully imagine now. 
The ancillary data sets in A2744 will be further enriched in upcoming months and years with the scheduled 
NIRCam/Wide-Field-Slitless-Spectroscopy (WFSS) observations (\#2883; PI F.~Sun, \#3561; PIs J.~Matthee \& R.~Naidu, \#3538; PI E.~Iani), NIRCam medium-band observations (\#4111; PI K.~Suess), and high-resolution deep ALMA Band~6 imaging (\#2023.1.00626.S; PI V.~Kokorev) that will accelerate and broaden the array of the legacy science. Below, we describe some of the key legacy science cases enabled by the synergy of ALMA and \jwst\ in A2744.

\subsection{What triggers dusty star formation$?$}
Even beyond the 69 ALMA-detected sources, the homogeneous ALMA 1.2-mm map automatically generates the best control sample of ALMA non-detected galaxies that also receive the same benefits from the rich \jwst\ data and the lensing magnifications.
This allows us to measure correlations between the 1.2-mm flux density (+upper limit) and rest-UV to optical properties such as morphology (e.g., merger, clumpiness), size, S\'ersic index, spatial offsets among emission, color gradient, the Balmer decrement, stellar mass, stellar age, metallicity that are all decisively probed by the deep sensitivity at 1--5$\mu$m with the improved spatial resolution ({$>\times4-10$ than {\it Hubble} and {\it Spitzer}}) owing to the deep NIRCam imaging plus lensing. 
The comparison of those physical properties between ALMA-detected and non-detected samples will allow us to investigate what are the key parameters regulating dust emission. 

\subsection{Search for ultra-high-$z$ galaxies}
\jwst\ has sparked a revolution of effort to discover and study galaxies at very early cosmic epochs. Dozens of high-redshift galaxy candidates have been identified at $z\simeq$ 9--17 towards both lensing clusters and blank fields \citep[e.g.,][]{atek2022, atek2023, bouwens2022c, bradley2022, castellano2022, donnan2023, labbe2023a, finkelstein2023, austin2023, leung2023, casey2023b}. Including some systematic spectroscopic measurements \citep[e.g.,][]{harikane2023b, fujimoto2023c}, their abundance at the bright-end ($M_{\rm UV} \lesssim -20$) exceeds nearly all theoretical predictions so far \citep[e.g.,][]{behroozi2015, behroozi2020, dayal2017, yung2019a, yung2020b, wilkins2022a, wilkins2022b,  mason2022, mauerhofer2023}, raising a tension even with the $\Lambda$CDM model in some cases \citep[e.g.,][]{bolyan-kolchin2023}. 
In this context, a noteworthy result is that a remarkably UV-bright galaxy candidate at $z\sim16$ turns out to be a lower-redshift galaxy at $z=4.9$ due to the underlying red continuum plus the strong optical line contributions to the NIRCam broad band photometry which mimics the high-redshift Ly$\alpha$ break feature (\citealt{arrabal-halo2023a}; see also e.g., \citealt{naidu2022b, zavala2023, fujimoto2022b, mackinney2023}). 
Although non-detection of the dust continuum in the ALMA map does not completely rule out the lower-$z$ solution in those ultra-high-$z$ candidates, it still rules out the possibility of the lower-$z$ dusty star-forming galaxy with SFR $\gtrsim30$~$M_{\odot}$~yr$^{-1}$ \citep{fujimoto2022b}. 
The presence of the homogeneous ALMA data in \survey\ is helpful in investigating the lower-$z$ possibility for (ultra-) high-$z$ candidates in A2744.

\subsection{Size and Morphology from UV, optical, to FIR}
\label{sec:size}

The galaxy size is one of the fundamental observables to quantify galaxy evolution, which is directly related to the mass assembly through star-forming activities. 
Therefore, the size and morphological studies in multi-wavelengths from rest-frame UV (un-obscured star-formation), optical (stellar), and FIR emission (obscured star-formation) are fundamental probes for the galaxy evolution in a comprehensive manner. 
Recent ALMA observations towards the classical SMG-class bright dusty starbursts ($S_{\rm 1mm}\gtrsim$ a few mJy) have revealed the compact sizes ($\lesssim1$-2~kpc; e.g.,  \citealt{ikarashi2015, simpson2015a, hodge2016, fujimoto2017, fujimoto2018, tadaki2020}) of their dusty star-forming regions that are comparable to that of the stellar distribution of the compact quiescent galaxies \citep[e.g.,][]{barro2013, williams2014, vanderwel2014}. 
These results start to connect the evolutionary sequence between the specific galaxy populations, while moderate star-forming activities occurring in a large variety of distant star-forming galaxies would also play an essential role in the mass assembly and, subsequently, the morphological transformation, given their long depletion time scales of $\sim0.5$--1~Gyr \citep[e.g.,][]{tacconi2010}
The ALMA sources identified in \survey\ are much fainter than the classical SMGs (Section~\ref{sec:phy_properties}), making them the optimal samples to study the role of the moderate star-forming activity. They are also helpful in studying their contribution to the mass assembly and the morphological transitions in a comprehensive manner from rest-frame UV, optical, and FIR wavelengths, which are leveraged by the gravitational lensing and the high-resolution images of \hst\ and \jwst.
Although the initial visual characterizations with NIRCam are presented in Section~\ref{sec:morphology}, 
the systematic size measurements, the morphology classification, and the presence or absence of the substructures (e.g., spiral arm, bar, star-forming clumps in the disk) in these multi-wavelengths will also be further investigated in a separate paper.

\subsection{Spatially-resolved Galaxy SEDs}
\label{sec:resolved}

A total of 20 NIRCam broad and medium-band filters, after completing the NIRCam Medium-band program (\#4111; PI K.~Suess), will offer an unprecedented opportunity to characterize the SEDs of distant galaxies, securely disentangling the stellar continuum and nebular emission lines. 
Furthermore, the SED analysis can be performed in spatially-resolved manners (e.g., pixel-to-pixel basis) owing to the high-resolution and great sensitivity of NIRCam, allowing us to investigate the spatial variations of the physical properties and resolve several problems in the spatially-integrated SED analysis such as the ``outshining'' from the recent bursts \citep[e.g.,][]{papovich2001, pforr2012, clara2022b, narayanan2023} and the break of the energy balance of the dust attenuation/re-emission due to the spatial offset between the dust-obscured and un-obscured regions directly observed with ALMA \citep[e.g.,][]{popping2017, narayanan2018, ferrara2022, kokorev2023}. The dedicated SED analysis will also constrain the SF history of the galaxies, and together with the spatially-resolved analysis, it will be an excellent probe for understanding the dramatic morphological transformations of the galaxies across cosmic time and how and where the first SF activities emerge. 
The high-resolution ALMA Band~6 imaging (\#2023.1.00626.S; PI V.~Kokorev) will also provide unique opportunities to directly investigate the associations of the dust-emitting regions and the underlying NIR properties.  

\subsection{Faint quasars/AGNs}
Recent \hst\ and \jwst\ observations routinely identify red compact sources at $z\sim3$--8 that are likely explained either by the faint quasar/AGN populations or compact dusty starbursts \citep[e.g.,][]{morishita2020,fujimoto2022, onoue2022, furtak2022b, endsley2023, furtak2023b, labbe2023, akins2023, barro2023, greene2023}. Despite their original small survey volumes, some of them have been already confirmed to be the quasar/AGN populations from the broad-line H$\beta$ detection with \jwst\ spectroscopy \citep{kocevski2023, matthee2023, furtak2023c, kokorev2023b, fujimoto2023c}, indicating that the faint end of the high-redshift quasar/AGN LFs is steeper than ever thought \citep[e.g.,][]{harikane2023c, maiolino2023c} based on the extrapolation of previous UV-optical-based type-I quasar/AGN studies \citep[e.g.,][]{matsuoka2018}, but likely close to the X-ray based quasar/AGN studies \citep[e.g.,][]{giallongo2019}. 
\cite{labbe2023} identify 26 red compact sources at $z\sim3$--7 in UNCOVER, where the dust continuum is not detected from any of them, and the joint \jwst+ALMA SED analysis prefers the models with the AGN-induced host dust rather than the dusty obscured star formation. 
Although the follow-up spectroscopy with the high spectral resolution is required to conclude whether these abundant red compact objects are faint AGNs (via e.g., identifications of broad lines and/or high ionization state lines) or other populations, 
the ALMA submm/mm observations will be a helpful probe for the red compact objects to disentangle their scenarios between the quasar/AGN or the compact dusty starburst statistically.  
It is worth mentioning that recent ALMA and deep X-ray studies also suggest a very high AGN fraction ($\sim90$\%) among the ALMA faint-mm sources at $z\sim2$ \citep{ueda2018}. In fact, some of our ALMA sources also show point-like morphology in the F444W band (Section~\ref{sec:morphology}). In addition to the red compact objects identified from the NIRCam data, the ALMA-detected sources are also exciting targets to understand the emergence of the faint AGNs at high redshifts.  

\subsection{Dust attenuation}
\label{sec:dust}
Understanding dust attenuation in galaxies is crucial to studying the true picture of the galaxies at all redshifts \citep[e.g.,][]{salim2020}.  
The dust attenuation will be securely measured via the Balmer decrement from the latest NIRSpec/MSA (Section~\ref{sec:nirspec}) and NIRCam/WFSS (\#2883, \#3561, \#3538) observations, which may be even achieved in spatially-resolved fashions (see Section~\ref{sec:nirspec}). 
Owing to the total of 20 NIRCam broad and medium-band filters, which securely separate the line and underlying continuum in each filter, 
the dust attenuation distribution will also be studied from the pixel-by-pixel-based SED analysis (Section~\ref{sec:resolved}). 
For the ALMA-detected sources, we can independently infer the total energy of the re-emitted thermal IR emission and directly measure the spatial positions of the dust emission. 
These indicate that comprehensive studies will be available from the dust obscuration to its re-emission in great detail.  
With the high-resolution NIRCam images, the secure measurements on the dust attenuation will also answer the question of whether it is dependent on the galaxy inclination \citep[e.g.,][]{wang2018, nelson2023, lorenz2023, gomez2023}. 

\subsection{SFR--$M_{\rm star}$--Metallicity relation}
\label{sec:metallicity}

Compared to the classical SMGs ($S_{\rm 1mm}\gtrsim$ a few mJy) that are dusty starburst galaxies with vigorously high SFR ($\gtrsim500$--1000~$M_{\odot}$~yr$^{-1}$), the faint submm/mm sources that are newly identified with ALMA start to capture the moderate star formation in the ordered disk or compact core \citep[e.g.,][]{rujopakarn2019, tadaki2020} and reveal that the majority ($>90$\%) of these faint ALMA sources fall within or even below the main-sequence of the SFR--$M_{\rm star}$ relation \citep[e.g.,][]{aravena2020}. To understand the evolutionary context of these faint ALMA sources, it is also important to study their chemical enrichment and investigate their distribution on the fundamental SFR--$M_{\rm star}$--Metallicity relation \citep[e.g.,][]{ucci2023}, where the gas-phase metallicity measurements via the optical line methods \citep[e.g.,][]{pettini2004} will be available with the scheduled NIRSpec/MSA (\#2561) and NIRCam/WFSS (\#2883, \#3561, \#3538) observations. 
The SFR and $M_{\rm star}$ parameters will also be measured with unprecedented levels owing to the total of 20 NIRCam filters both from all available broad and medium-band filters, which the secure dust correction will also refine via the Balmer decrement  (Section~\ref{sec:dust}) and the spatially-resolved SED fitting (Section~\ref{sec:resolved}).

\subsection{1.2-mm Number counts}
Recent ALMA observations allow us to explore a faint submm/mm regime ($S_{\rm 1mm}<1$ mJy) without uncertainties from source confusion and blending, owing to ALMA's high sensitivity and angular resolution relative to the single-dish telescopes. 
However, the survey area in the deepest layer ($S_{\rm 1mm}\simeq0.01$--0.1~mJy) is still very limited due to its small field of view (see Figure~\ref{fig:survey_area}), and thus there remains large uncertainty in the faint-end slope estimate in the submm/mm number counts \citep[e.g.,][]{gonzalez2020, fujimoto2023b}. 
As a result, the origin of the Cosmic Infrared Background light (CIB) has not yet been fully accounted for as yet, despite its importance implied by the fact that the total energy of the CIB has been known to be comparable to the Cosmic Optical Background light since its initial discovery with the {\it Cosmic Background Explorer} satellite \citep{puget1996,fixsen1998,hauser1998,hauser2001,dole2006}.  
Given the deepest survey area newly added by \survey\ (see Figure~\ref{fig:survey_area}) and the 69 ALMA sources are already identified with high purity (Section~\ref{sec:source_ext}), \survey\ provides essential constraints on the faint-end of the 1.2-mm number counts independent from previous surveys. Although the magnification uncertainty could be a complication in these lensing studies, the secure $z_{\rm phot}$ measurements with the rich NIRCam data sets are beneficial to mitigate the magnification uncertainty, which will be further refined with the scheduled NIRCam medium-filter observations (\#4111) and the NIRSpec/MSA (\#2561) and NIRCam/WFSS (\#2883, \#3561, \# 3538) spectroscopy. 

\subsection{IRLF \& Cosmic SFRD at $z\gtrsim3$}
While we show the initial results of the IRLF measurement out to $z\sim10$ in Section~\ref{sec:irlf}, a complete measurement will be further required, including the completeness correction based on the proper size measurements and realistic uncertainties in the 1.2-mm flux and magnification estimates. For the \jwst-dark ALMA galaxy candidates and the marginal ALMA detection from the X-ray AGN at $z\sim10$, the confirmations of their ALMA detection and spectroscopic redshifts are essential in the first place. These results will provide important constraints on the obscured side of the star-forming activity in the universe at $z\gtrsim3$ that has not yet been well constrained with the blind submm/mm surveys so far. 

\subsection{CO, \ci, and \cii\ line LFs}
The cold interstellar medium (ISM), such as the neutral atomic gas and the dense molecular gas, are key elements regulating the galaxy formation and evolution as the fuel of the star formation, and thus CO, \ci, and \cii\ LF measurements and their evolution are important probes to understand the cosmic SFR history. 
With the lensing support, the 30-GHz-wide ($\sim244$--274~GHz) ALMA cube of \survey\ will be a powerful probe for the medium-$J$ CO transitions ($4\lesssim J_{\rm up}\lesssim8$) and two \ci\ transitions at $z\sim1$--3 and $z\sim6$--7 \cii\ line. 
The rich NIRCam data sets are very helpful to nail down the redshift solution even when only a single FIR line is detected in the ALMA data cube. 
In the deep ALMA Band~6 observations in ASPECS \citep{walter2018}, 35 moderately secure line emitters have been identified \citep{decarli2020}. Given the larger survey area of \survey\ than that of ASPECS (see Figure~\ref{fig:survey_area}), resulting in $\sim3$ times more continuum source identification in \survey\footnote{25 sources are identified in the ASPECS Band~6 observations with the same SNR threshold at $\sim5.0$ \citep{gonzalez2020}.}, a simple scaling suggests that we may expect to identify $\sim100$ FIR line emitters in the \survey\ data cube that are similarly secure to the FIR line emitters identified in ASPECS. 
These indicate that the ALMA data cube of \survey\ is one of the best data sets to conduct FIR line LF measurements beyond the lower limit in the $z=6$ \cii\ LF from the successful identification of the single \cii\ line emitter at $z=6.33$ (Section~\ref{sec:cii_emitter}). 

\subsection{Faint-end of SFR-$L_{\rm [CII]}$ relation at $z=6-7$}
The 30-GHz frequency setup covers the \cii\ emission at $z=$5.94--6.79. 
In addition to one successful \cii\ line identification demonstrated in Section~\ref{sec:cii_emitter}, a complete search may identify more \cii\ line emitters within the above redshift range. 
Apart from the blind search, the scheduled multiple \jwst\ spectroscopic programs using NIRCam/WFSS and NIRSpec/MSA will surely dramatically increase the spec-$z$ sample via the rest-frame optical emission lines within the above redshift range. Thus the \cii-line stacking with the ALMA data cube will also be available without concerns of the velocity offset of Ly$\alpha$ \citep[e.g.,][]{jolly2021}. 
If we assume $N=100$, a typical magnification of $\mu=2$, and a line width of 150 km~s$^{-1}$, the stacked \cii\ spectrum reaches sensitivity down to $L_{\rm [CII]}\simeq$ $1\times10^{7}\,L_{\odot}$, which corresponds to SFR$\sim$1 $M_{\odot}$~yr$^{-1}$ based on the SFR--$L_{\rm [CII]}$ relation calibrated among the local galaxies (\citealt{delooze2014}; see also e.g., \citealt{leung2020, liang2023}). This level of the faint-end of SFR--$L_{\rm [CII]}$ relation has never been explored even in recent ALMA large programs (e.g., ASPECS, ALPINE, ALCS), and providing important insights on the cold ISM properties of the low-mass early galaxies. Such constraints are also beneficial for predictions on the future \cii\ intensity mapping experiment \citep[e.g.,][]{yue2019,syang2022}.  

\section{Summary}
\label{sec:summary}
In this paper, we present ALMA Band~6 observations of \survey\, which is designed to establish the first joint ALMA and \jwst\ public legacy field. 
The ALMA observations achieve a homogeneous 1.2-mm mosaic mapping towards the massive galaxy cluster A2744 over a $4'\times6'$ area that has also been observed in a deep and homogeneous NIRCam+NIRSpec program of the \jwst\ treasury program UNCOVER \citep{bezanson2022}, where many more imaging and spectroscopic programs are scheduled in \jwst\ Cycle~2. 
The multiple frequency setups are used for the ALMA observations to continuously cover the 244--274~GHz range, which maximizes the identification of the line emitters as well as the continuum sources in a blind manner. 
The major findings of this paper are summarized below: 
\begin{enumerate}
\item The observations achieved the continuum sensitivity down to $\sigma=$32.7~$\mu$Jy over the $4'\times6'$ area, homogeneously enlarging the ALMA survey area around the A2744 by $\sim6$ times more than the previous ALMA programs in this field. We produced the wide homogeneous and deep maps by combining the previous data around the primary cluster region, and identified 69 continuum sources with peak pixel SNR $\gtrsim$ 5.0. The positive and negative source analysis suggests that there may be one or two spurious sources above our SNR thresholds, yielding the purity among our 69 continuum sources of $>0.97$.  
\item Out of the 69 ALMA continuum sources, 67 sources have counterparts in the deep NIRCam maps with the spatial offset of $0\farcs0$--$0\farcs70$, equal to $0$--30\% probability of the chance projection based on the surface density of the NIRCam sources. 
Seventeen ALMA sources have been observed in the NIRSpec prism follow-up with MSA, where multiple emission lines are successfully detected from all MSA-observed ALMA sources, securely determining their source redshifts. The NIRSpec prism spectra taken in different shutters of MSA show the spatial variation of the dust attenuation via Pa$\gamma$/Pa$\beta$ in an ALMA-detected galaxy at $z=2.985$, demonstrating the power of the joint ALMA and NIRSpec MSA analysis to gain insights into dust-obscured properties in high-redshift galaxies, also in a spatially-resolved manner. 
\item Leveraged by the latest NIRSpec prism spectroscopy, a total of 27 ALMA sources have the spectroscopic redshifts, while the photometric redshifts are also constrained with \prospector\ and \eazy\ codes using the comprehensive \hst, \jwst, and ALMA data sets for the remaining sources. These ALMA sources show the redshifts (the median values) of $z=0.29$--9.89 (2.30), the lensing magnifications of $\mu=$1.0--9.3 (1.8), and the intrinsic 1.2-mm flux densities after the lensing correction of $S_{\rm 1.2mm}^{\rm int}=$ 0.04--1.65~mJy (0.24~mJy).  
\item Almost all NIRCam counterparts show undisturbed, ordered morphologies either by disks or spheroids. Although some have potential merging companions nearby, the dust continuum arises not from the potential merging plane but around the central region of the counterpart, indicating a low merging fraction ($<10$\%) for the ALMA continuum sources in UNCOVER.  
In contrast to the disturbed morphology observed in the majority ($\sim80\%$) of the bright submm galaxies (SMGs; $>$ a few mJy at submm/mm), this indicates that the faint ALMA mm sources display less violent mechanisms than merging events. 
\item By using the color and magnitude criteria of F150W$-$F444W$>$2.3 and F150W$>$27.0~mag, we identify eight \hst-dark galaxies among the ALMA continuum sources that are characterized with $z=2.58$--4.79 and $\log(M_{\rm star}/M_{\odot})=9.81$--10.66 by the \prospector\ fit.  
From the NIRCam maps, the dust lane is clearly observed in some of these \hst-dark ALMA galaxies with edge morphology, exactly from where the dust emission arises. However, several \hst-dark ALMA galaxies show face-on morphology in contrast, indicating that the inclination does not always cause significant dust obscuration. 
\item We also identify two candidates of the \jwst-dark galaxy among our ALMA sources that are fainter than 30.0~mag in F444W after the lens correction. Several potential counterparts are visible in the F277W+F356W+F444W detection map, where one of the possible counterparts with a spatial offset of $0\farcs25$ shows the photometric redshift of $z\sim9$ by our SED fits. Although there still remains the possibility that these ALMA sources are spurious, the probability of the chance projection of the $z\sim9$ NIRCam faint source is estimated to be $\sim0.04\%$. 
\item By analyzing 30-GHz-wide Band~6 spectra extracted at 150 bright (F150W$<$27.5~mag) NIRCam source positions whose redshift estimates are $z\sim6-7$, we identify one secure (SNR $=$ 7.0) emission line at $259.445\pm0.015$~GHz with a line width of FWHM $=220\pm40$~km~s$^{-1}$. This corresponds to the \cii\ redshift at $z=6.3254\pm0.0004$, which is spectroscopically confirmed in the follow-up NIRSpec spectroscopy. The key optical emission lines are all detected, such as H$\alpha$+\nii, \oiii5008, H$\beta$, H$\gamma$, \oii3727+3730, and \neiii3869, With the strong line calibrations, the gas-phase metallicity is securely measured to be 12+$\log$(O/H) = $7.84^{+0.25}_{-0.16}$, where the observed \cii\ luminosity is consistent with the typical SFR-$L_{\rm [CII]}$ relation both from observations and theoretical models within the errors. 
This successful \cii\ line identification provides a lower limit of $3.2\times10^{-5}$~Mpc$^{-3}$ in the \cii\ luminosity function (LF) at $z\sim6$ at $\log(L_{\rm [CII]}/L_{\odot})=8.8$, which places higher than predictions from semi-analytical models. 
\item By searching for any marginal detection from 16 spectroscopic and photometric galaxies at $z\gtrsim9$ presented in \cite{fujimoto2023c} and \cite{atek2023}, we find that none of these candidates show a dust continuum emission above $3\sigma$ levels, while a marginal continuum detection ($2.6\sigma$) takes place only at the X-ray-detected AGN host galaxy at $z=10.07$. While this marginal ALMA continuum could be spurious, the sheer coincidence of such a marginal detection identified only from the X-ray AGN host galaxy at $z>10$ may indicate an active co-evolution of the early massive black hole and its host. 
\item Based on the secure redshift constraints for our ALMA sources, we derive the infrared (IR) LFs at $z=1-5$. We find that our IRLF measurements are consistent with previous results, but likely have a little underestimate of the faint end, which is explained by the lack of completeness correction in our measurements. 
By assuming that the ALMA emission is all real for the \jwst-dark galaxy candidates and the marginal detection in the X-ray AGN candidate, we also derive possible constraints in the IRLF at $z\sim10$. The possible constraints are consistent with predictions from the galaxy formation models, indicating that identifying the \jwst-dark galaxy candidates and the faint mm emission from the X-ray AGN host galaxy at $z=10.07$ is not unfeasible in the abundance perspective.  
\item We also introduce several key legacy science cases that are enabled by the synergy of ALMA and \jwst\ in A2744, including NIRCam/Wide-Field-Slitless-Spectroscopy, NIRCam medium-band imaging observations, and high-resolution ALMA imaging that are all scheduled in A2744 in upcoming months and years.   
\end{enumerate}

We thank the anonymous referee for the careful review and valuable comments that improved the clarity of the paper. 
We thank Caitlin Casey, Jorge Zavala, and Arianna Long for discussing the IRLF measurements, and Ian Smail for useful comments on the paper.   
This paper makes use of the ALMA data: ADS/JAO. ALMA \#2022.1.00073.S, 2018.1.00035.L, and 2013.1.00999.S. 
ALMA is a partnership of the ESO (representing its member states), 
NSF (USA) and NINS (Japan), together with NRC (Canada), MOST and ASIAA (Taiwan), and KASI (Republic of Korea), 
in cooperation with the Republic of Chile. 
The Joint ALMA Observatory is operated by the ESO, AUI/NRAO, and NAOJ. 
This work is based on observations and archival data made with the {\it Spitzer Space Telescope}, which is operated by the Jet Propulsion
Laboratory, California Institute of Technology, under a contract with NASA along with archival data from the NASA/ESA 
{\it Hubble Space Telescope}. 
This project has received funding from NASA through the NASA Hubble Fellowship grant HST-HF2-51505.001-A awarded by the Space Telescope Science Institute, which is operated by the Association of Universities for Research in Astronomy, Incorporated, under NASA contract NAS5-26555.
PD acknowledges support from the NWO grant 016.VIDI.189.162 (``ODIN") and from the European Commission's and University of Groningen's CO-FUND Rosalind Franklin program.
AZ and LJF acknowledge support by Grant No. 2020750 from the United States-Israel Binational Science Foundation (BSF) and Grant No. 2109066 from the United States National Science Foundation (NSF), and by the Ministry of Science \& Technology, Israel. IL acknowledges support by the Australian Research Council through Future Fellowship FT220100798.
The project leading to this publication has received support from ORP, that is funded by the European Union’s Horizon 2020 research and innovation programme under grant agreement No 101004719 [ORP]. 

Some/all of the data presented in this paper were obtained from the Mikulski Archive for Space Telescopes (MAST) at the Space Telescope Science Institute via \dataset[doi:10.17909/nftp-e621]{http://dx.doi.org/10.17909/nftp-e621} and
\dataset[doi:10.17909/8k5c-xr27]{http://dx.doi.org/10.17909/8k5c-xr27]}
The reduced \jwst/NIRCam data (DR2) is all available via \url{https://jwst-uncover.github.io/}.
For ALMA, the original raw data can be accessed via \url{https://almascience.nao.ac.jp/aq/}. 
The reduced ALMA continuum maps and cubes are also available via \url{https://jwst-uncover.github.io/DR2.html#DUALZ}. 

\software{{\sc casa} (v6.4.1; \citealt{casa2022}), {\sc Source Extractor} \citep{bertin1996},  {\sc astropy} \citep{astropy2013}.
}

\clearpage

\appendix

\bibliographystyle{apj}
\bibliography{apj-jour,reference}

\end{document}